\documentclass[a4paper,11pt]{article}
\usepackage{jheppub}

\usepackage{amsfonts}
\usepackage{mathrsfs}
\usepackage{graphicx}
\usepackage{amsmath}
\usepackage{amssymb}

\usepackage{epsfig}
\usepackage{slashed}
\usepackage{color}
\usepackage{multirow}
\usepackage{ulem}
\usepackage{adjustbox}
\usepackage{stmaryrd}
\usepackage{pdflscape}
\usepackage[figuresright]{rotating}

\newcommand{\bqa}{\begin{eqnarray}}
	\newcommand{\eqa}{\end{eqnarray}}
\newcommand{\beq}{\begin{equation}}
	\newcommand{\eeq}{\end{equation}}
\def\non{\nonumber}

\allowdisplaybreaks[1]

\graphicspath{{fig/}{dia/}}
\DeclareGraphicsExtensions{.eps}

\title{Study of doubly heavy baryon lifetimes}

\author[a]{Hai-Yang Cheng,}
\author[b]{Chia-Wei Liu}

\affiliation[a]{Institute of Physics, Academia Sinica, Taipei 115, Taiwan}
\affiliation[b]{School of Fundamental Physics and Mathematical Sciences, Hangzhou Institute for Advanced Study, UCAS, Hangzhou 310024, China}

\emailAdd{phcheng@phys.sinica.edu.tw}
\emailAdd{chiaweiliu@ucas.ac.cn}

\abstract{
We study the lifetimes and inclusive semileptonic decay widths of doubly heavy baryons within the framework of heavy quark expansion. Our analysis includes next-to-leading-order corrections to the dimension-3, -5, and -6 operators, together with the leading dimension-7 contributions, while the nonperturbative matrix elements are evaluated in a bag model with translationally improved baryon wave functions. We obtain $( \tau_{\Xi_{cc}^{++}} , \tau_{\Xi_{cc}^{+}} , \tau_{\Omega_{cc}^{+}} ) = ( 2.67^{+0.95}_{-0.94},\, 0.47^{+0.09}_{-0.08},\, 1.79^{+0.64}_{-0.62} ) \times 10^{-13}\,{\rm s}$ and $( \tau_{\Xi_{bb}^{0}} , \tau_{\Xi_{bb}^{-}} , \tau_{\Omega_{bb}^{-}} ) = ( 0.75^{+0.13}_{-0.11},\, 0.92^{+0.15}_{-0.15},\, 0.93^{+0.15}_{-0.15} ) \times 10^{-12}\,{\rm s}$.
The lifetime hierarchy patterns are found to be  $\tau(\Xi_{cc}^{++})>\tau(\Omega_{cc}^+)>\tau(\Xi_{cc}^+)$
and $\tau(\Omega_{bb}^{-})\sim\tau(\Xi_{bb}^-)>\tau(\Xi_{bb}^0)$ for doubly charmed and bottom baryons, respectively. 
The $W$-exchange contribution plays a crucial role in generating the large lifetime splitting in the doubly charmed sector and remains phenomenologically important for doubly bottom baryons.
{In particular, for $\Xi_{cc}^{+}$ the dominant dimension-6 spectator effect, namely $W$-exchange, surpasses the nominally leading term. The same spectator operator is realized as destructive Pauli interference in $D^+$, owing to the $d$ versus $\bar d$ spectator. Although the $\Xi_{cc}^{+}$ case does not lead to the negative decay width encountered for $D^+$, the absolute value of $\tau(\Xi_{cc}^{+})$ should nevertheless be regarded as only indicative.} 
 In addition to the total lifetimes, we calculate the separate nonleptonic and semileptonic contributions, which allow us to trace the pattern of spectator effects in each baryon channel. We also evaluate the inclusive semileptonic decay widths and the decay width asymmetries, which provide complementary probes of the underlying decay mechanisms. 
}

\begin{document}
	\maketitle
	
	\section{Introduction}
	The interest in the lifetimes of charmed baryons has been intensively revived both experimentally and theoretically since 2018. Back to 1986 the lifetime pattern of charmed baryons had already been predicted to be
	\cite{Shifman:1986mx,Guberina:1986gd}
	\begin{equation} \label{eq:Bclifetimehierarchy}
		\tau(\Xi_c^+)>\tau(\Lambda_c^+)>\tau(\Xi_c^0)>\tau(\Omega_c^0).
	\end{equation}
	According to the 2004 version of Particle Data Group (PDG) \cite{PDG2004}, the world averages of their lifetimes then were given by
	\begin{eqnarray} \label{eq:exptlifetime}
		&& \tau(\Lambda^+_c)= (200\pm 6)\,{\rm fs}, \quad
		\tau(\Xi^+_c)= (442\pm 26)\,{\rm fs},
		\nonumber \\
		&& \tau(\Xi^0_c)= (112^{+13}_{-10})\,{\rm fs}, \quad~~
		\tau(\Omega^0_c)= (69\pm12)\,{\rm fs}.
	\end{eqnarray}
	These world averages  remained stable from 2004 till 2018 \cite{PDG2018}. The predicted lifetime hierarchy for charmed baryons given in eq. (\ref{eq:Bclifetimehierarchy}) based on the framework of heavy quark expansion (HQE) thus agrees with experimental measurements given in eq. (\ref{eq:exptlifetime}).
	
	However, the situation was dramatically changed in 2018 when LHCb reported a new measurement of the charmed baryon $\Omega_c^0$ lifetime using semileptonic $b$-hadron decays \cite{LHCb:2018nfa}.  LHCb found $\tau(\Omega_c^0)=(268\pm24\pm10\pm2)\,{\rm fs}$, using the semileptonic $\Omega_b^-$ decay. This value is nearly four times larger than the 2018 world-average value of $\tau(\Omega_c^0)$ (see eq. (\ref{eq:exptlifetime})) extracted from fixed target experiments. As a result, a new lifetime pattern emerged
	\begin{equation} \label{eq:newhierarchy}
		\tau(\Xi_c^+)>{\tau(\Omega_c^0)}>\tau(\Lambda_c^+)>\tau(\Xi_c^0).
	\end{equation}
	This measurement of the $\Omega_c^0$ lifetime was subsequently confirmed by LHCb in 2021 using promptly produced $\Omega_c^0$ baryons \cite{LHCb:2021vll}, and again in 2025 using hadronic $b$-hadron decays \cite{LHCb:2025oww}.
	Belle has also performed a new measurement of $\tau(\Omega_c^0)$ in 2022 \cite{Belle-II:2022plj}. Combining the current PDG values \cite{PDG2024} with the new LHCb results for the lifetimes of $\Xi_c^0$ and $\Omega_c^0$ \cite{LHCb:2025oww}, the new world averages now read
	\begin{eqnarray} \label{eq:newexptlifetime}
		&& \tau(\Lambda^+_c)= (202.6\pm 1.0)\,{\rm fs}, \quad
		\tau(\Xi^+_c)= (453\pm 5)\,{\rm fs},
		\nonumber \\
		&& \tau(\Xi^0_c)= (149.8\pm1.9)\,{\rm fs}, \quad~
		\tau(\Omega^0_c)= (274\pm10)\,{\rm fs}.
	\end{eqnarray}
	
	In view of the very striking and astonishing observation of a huge jump of the $\Omega_c^0$ lifetime, it is natural to raise the following two questions: (i) Is it possible to have the large $\Omega_c^0$ lifetime measured much earlier than 2018? (ii) How to understand the new hierarchy of charmed baryon lifetimes given in eq. (\ref{eq:newhierarchy})? The first question has been addressed by Bianco and Bigi \cite{Bianco:2020hzf}. They accentuated that the $\Omega_c^0$ lifetime value measured by LHCb is so large that could have been easily measured much earlier than 2018 by experiments at $e^+e^-$ colliders such as CLEO-c and Belle with resolution about 150 fs typically. As for the second question, dimension-7  four-quark operators associated with the $1/m_c^4$ term in HQE give leading corrections to the spectator effects and are responsible for the new hierarchy of charmed baryon lifetimes. For the sake of the ensuring discussions, the reader is referred to Table 6 of \cite{Cheng:2023jpz} (or Tables 24 and 25 of \cite{Gratrex:2022xpm}). We see from the table that in the HQE approach up to $1/m_c^3$ expansion, the $\Omega_c^0$ is shortest-lived as it receives largest constructive Pauli interference in both nonleptonic and semileptonic decays denoted by $\Gamma_6^{\rm NL}$ and $\Gamma_6^{\rm SL}$ from dimension-6 operators, respectively. This leads to the old lifetime hierarchy pattern $\tau(\Xi_c^+)>\tau(\Lambda_c^+)>\tau(\Xi_c^0)>\tau(\Omega_c^0)$. From the table we also see that $\Gamma_7^{\rm SL}$ from dimension-7 four-quark operators contributes destructively to the total semileptonic rate $\Gamma^{\rm SL}$, while $\Gamma_7^{\rm NL}$ contributes constructively to $\Gamma^{\rm NL}$ for $\Lambda_c^+$ and $\Xi_c^0$ and destructively for $\Omega_c^0$ and $\Xi_c^+$. Owing to the presence of two strange quarks in the $\Omega_c^0$, $\Gamma_7^{\rm NL}(\Omega_c^0)$ and $\Gamma_7^{\rm SL}(\Omega_c^0)$ are the largest in magnitude among the charmed baryons.
	Consequently, the old lifetime hierarchy pattern is modified to the new one $\tau(\Xi_c^+)>\tau(\Omega_c^0)>\tau(\Lambda_c^+)>\tau(\Xi_c^0)$. \footnote{The value of $\tau(\Omega_c^0)\sim 2.3\times 10^{-13}s$ and the possibility that $\Omega_c^0$ lives longer than $\Lambda_c^+$ were emphasized by one of us (H.Y.C.) in \cite{Cheng:HIEPA} even before the first LHCb measurement of the $\Omega_c^0$ lifetime. An analysis of the effects of dimension-7 four-quark operators on the charmed baryon lifetimes was performed in \cite{Cheng:2018rkz}.}

The standard theoretical framework for the study of heavy hadron lifetimes is based on HQE. 
The major uncertainties stem from the baryonic matrix elements which are customarily evaluated  using the non-relativistic quark model (NRQM). 
	For the singly heavy baryon, say $\Lambda_b$, the matrix element of four-quark operator $L^q_{\Lambda_b}$ to be defined below is related to $\Lambda_b$ wave function modulus squared at the origin; that is $L^q_{\Lambda_b}=-|\psi_{bq}^{\Lambda_b}(0)|^2$, which is related to $|\psi_{bq}^B(0)|^2=\frac{1}{12}f_B^2 m_B$ through the hyperfine mass splitting in the bottom baryon and the $B$ meson sectors. In the heavy quark limit, $f_B^2\propto m_b^{-1}$ and $f_D^2\propto m_c^{-1}$. Hence, the relation $L_{\Lambda_c}^q=L_{\Lambda_b}^q$ holds in the heavy quark limit. However, for finite quark masses, $f_B^2m_B\gg f_D^2m_D$ and $L_{\Lambda_b}^q\approx 2.5 L_{\Lambda_c}^q$ found in the realistic NRQM calculation, which is far from the naive expectation of the heavy quark limit \cite{Cheng:2023jpz}. Nevertheless, this difficulty can be circumvented in the bag model (BM) in which the baryon wave function is translationally improved. Indeed, it is evident from Table 5 of 
\cite{Cheng:2023jpz} that the matrix element $L_{{\cal B}_Q}$ evaluated in the BM varies less than 10\%  with respect to the heavy flavor. This is why it is more sensible to evaluate the baryonic matrix elements using the BM.
{As discussed below, however, this near stability for singly heavy baryons does not persist quantitatively in the doubly heavy baryons, where the spectrum-derived bag radii lead to a much larger heavy-flavor breaking.}
    
	In this work we shall focus on the lifetimes of doubly charmed baryons $\Xi_{cc}^{++}$, $\Xi_{cc}^+$, $\Omega_{cc}^+$ and doubly bottom baryons $\Xi_{bb}^0$, $\Xi_{bb}^-$ and $\Omega_{bb}^-$. The mass and the lifetime of $\Xi_{cc}^{++}$ have been measured by LHCb to be $3621.55\pm0.38$ MeV \cite{LHCb:2019epo} and $256\pm 27$ fs \cite{LHCb:2018zpl}, respectively. The $\Xi_{cc}^+$ state was recently observed by LHCb through its decay to the $\Lambda_c^+ K^-\pi^+$ final state \cite{LHCb:2026pxn}. Its mass was measured to be $3619.97\pm0.83\pm0.26^{+1.90}_{-1.30}$ MeV. Theoretical predictions of doubly charmed baryon lifetimes 
\cite{Kiselev:1999,Guberina,Kiselev:2002,Chang,Karliner:2014,Cheng:doubly,Berezhnoy,Dulibic:2023jeu}	
    are displayed in Table 1 (see also Fig. \ref{fig:lifetime_dcc} below). It turns out that both $\tau(\Xi_{cc}^{++})$ and
	$\tau(\Xi_{cc}^+)$ spread a large range. 
 As for the doubly bottom baryons, most of the calculations 
\cite{Likhoded:1999yv,Kiselev:2002,Chang,Karliner:2014,Berezhnoy} lead to 
$\tau(\Xi_{bb}^-)$ and $\tau(\Xi_{bb}^0)$
to be very similar, whereas the analysis in \cite{Cheng:2019sxr} indicates that $\tau(\Xi_{bb}^-)>\tau(\Xi_{bb}^0)$ as shown in Table~\ref{tab:lifetimes_db}.

In the present work, we shall follow our previous analysis \cite{Cheng:2023jpz} to evaluate the matrix elements in the BM to see if the 2-quark matrix elements $\mu_\pi^2$, $\mu_G^2$ and $\rho_D^3$ to be defined below are independent of the heavy quark mass $m_Q$ in the heavy quark limit, and so are the matrix elements   of the four-quark operator. We will also include
 next-to-leading-order~(NLO) corrections to the dimension-3, -5 and -6 operators in HQE to improve the predictions by one of us (H.Y.C.) in regard to the lifetimes of ${\cal B}_{cc}$ \cite{Cheng:doubly} and ${\cal B}_{bb}$~\cite{Cheng:2019sxr}.

This paper is organized as follows. In Sec. 2 we set up the framework and background of HQE necessary for the study of doubly heavy baryon lifetimes. Sec. 3 is devoted to evaluating the baryon matrix elements using the improved BM. Numerical results and discussions are presented in Sec. 4. We conclude this work in Sec. 5.
	
	\begin{table}[t]
		\caption{Predicted lifetimes of doubly charmed baryons
			in units of $10^{-13}s$. The results of \cite{Berezhnoy} are based on the calculation of using $m_c=1.73\pm0.07$ GeV and $m_s=0.35\pm0.20$ GeV from a fit to the LHCb measurement of $\tau(\Xi_{cc}^{++})$. The predicted lifetimes in  \cite{Dulibic:2023jeu} are calculated in the kinetic and pole mass schemes.
		} \label{tab:lifetimes_dc}
		\begin{center}
			\begin{tabular}{l c c c } \hline \hline
				& ~~$\Xi_{cc}^{++}$~~ & ~~$\Xi_{cc}^{+}$~~ & ~~$\Omega_{cc}^{+}$~~~~~\\
				\hline
				~~Kiselev et al.  \cite{Kiselev:1999}~~ &  ~~~$4.3\pm1.1$~~~ & ~~~$1.1\pm0.3$~~~ &  \\
				~~Guberina et al.  \cite{Guberina}~~ & 15.5 & 2.2 & 2.5 \\
				~~Kiselev et al.  \cite{Kiselev:2002}~~ & $4.6\pm0.5$ & $1.6\pm0.5$ & $2.7\pm0.6$ \\
				~~Chang et al.  \cite{Chang}~~  & 6.7 & 2.5 & 2.1 \\
				~~Karliner, Rosner  \cite{Karliner:2014}~~ & 1.85 & 0.53 & \\
				~~Cheng, Shi  \cite{Cheng:doubly} & 2.98 & 0.44 & 0.75 $\sim$ 1.80\\
				~~Berezhnoy et al.  \cite{Berezhnoy} & $2.6\pm0.3$ & $1.4\pm0.1$ & $1.8\pm0.2$ \\
				~~Dulibic et al. (Kinetic) \cite{Dulibic:2023jeu} & $3.2\pm0.5^{+0.8}_{-0.7}$ & $0.6\pm0.1^{+0.2}_{-0.1}$
				& $1.5\pm0.4^{+0.3}_{-0.2}$ \\
				~~Dulibic et al. (Pole) \cite{Dulibic:2023jeu} & $3.2\pm0.4^{+0.9}_{-0.7}$ & $0.6\pm0.1^{+0.2}_{-0.1}$
				& $1.6\pm0.5^{+0.4}_{-0.3}$ \\
				\hline
				~~Expt. \cite{LHCb:2018zpl} & $2.56^{+0.24}_{-0.22}\pm0.14$ & & \\
				\hline \hline
			\end{tabular}
		\end{center}
	\end{table}

	\begin{table}[t]
		\caption{Predicted lifetimes of doubly bottom baryons
			in units of $10^{-13}s$. 
		} \label{tab:lifetimes_db}
		\begin{center}
			\begin{tabular}{l c c c } \hline \hline
				& ~~$\Xi_{bb}^{0}$~~ & ~~$\Xi_{bb}^{-}$~~ & ~~$\Omega_{bb}^{-}$~~~~~\\
				\hline
				~~Likhoded et al.  \cite{Likhoded:1999yv}~~ &  7.9& 8.0 & 8.0  \\
				~~Kiselev et al.  \cite{Kiselev:2002}~~ & $7.9$ & $8.0$ & $8.0$ \\
				~~Karliner, Rosner  \cite{Karliner:2014}~~ & 3.7 & 3.7 & \\
				~~Berezhnoy et al.  \cite{Berezhnoy} & $5.2\pm0.095$ & $5.3\pm0.096$ & $5.3\pm0.093$ \\
                ~~Cheng, Xu  \cite{Cheng:2019sxr} & 6.87 & 8.65 & 8.68 \\
				\hline \hline
			\end{tabular}
		\end{center}
	\end{table}

	\section{Theoretical framework}\label{sec2}
	
\begin{figure}[t]
		\begin{center}
			\begin{minipage}{0.25\linewidth}
				\centering
				\includegraphics[width=\linewidth]{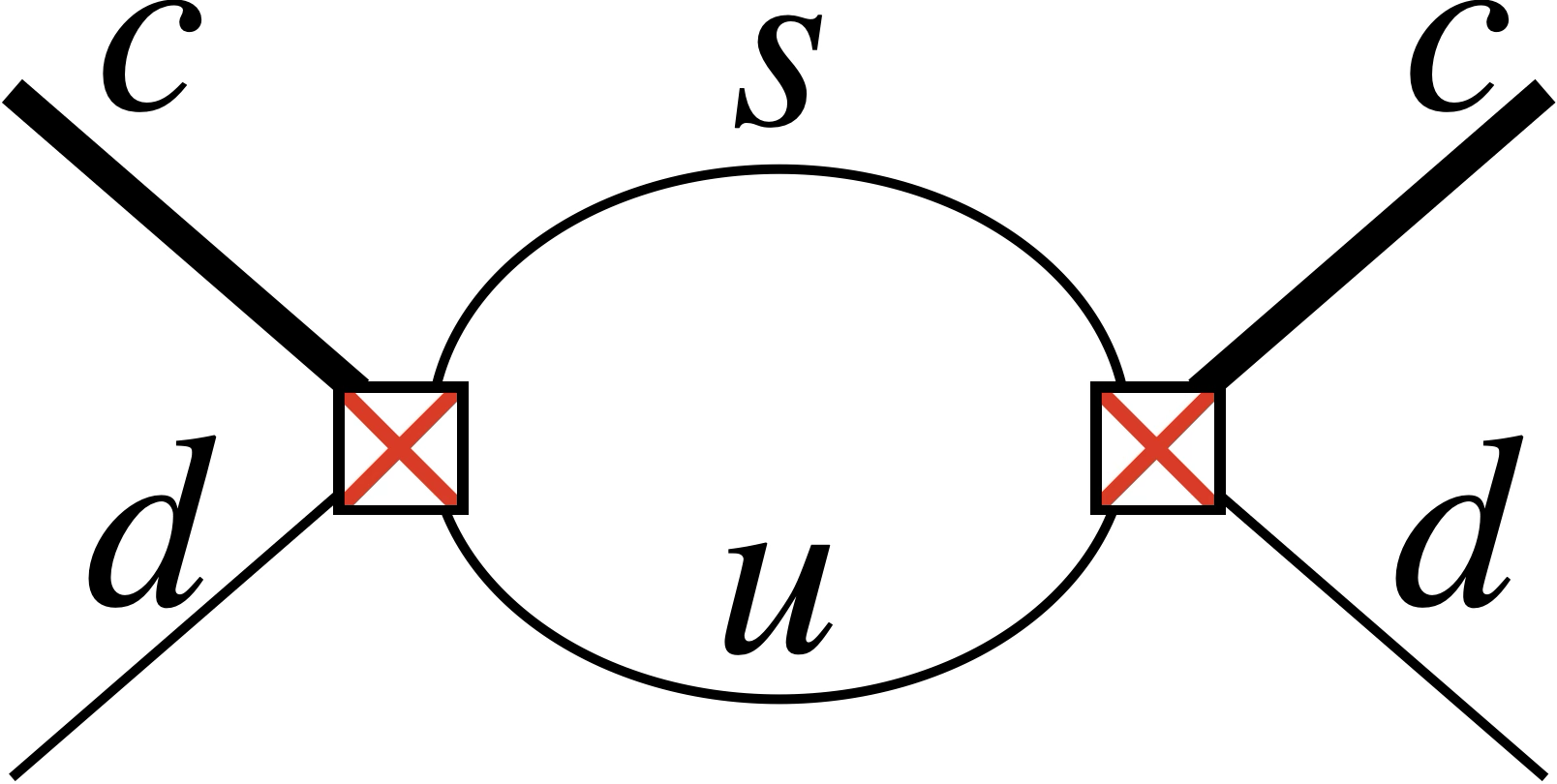}\\
				(a)
			\end{minipage}~~~~~
			\begin{minipage}{0.25\linewidth}
				\centering
				\includegraphics[width=\linewidth]{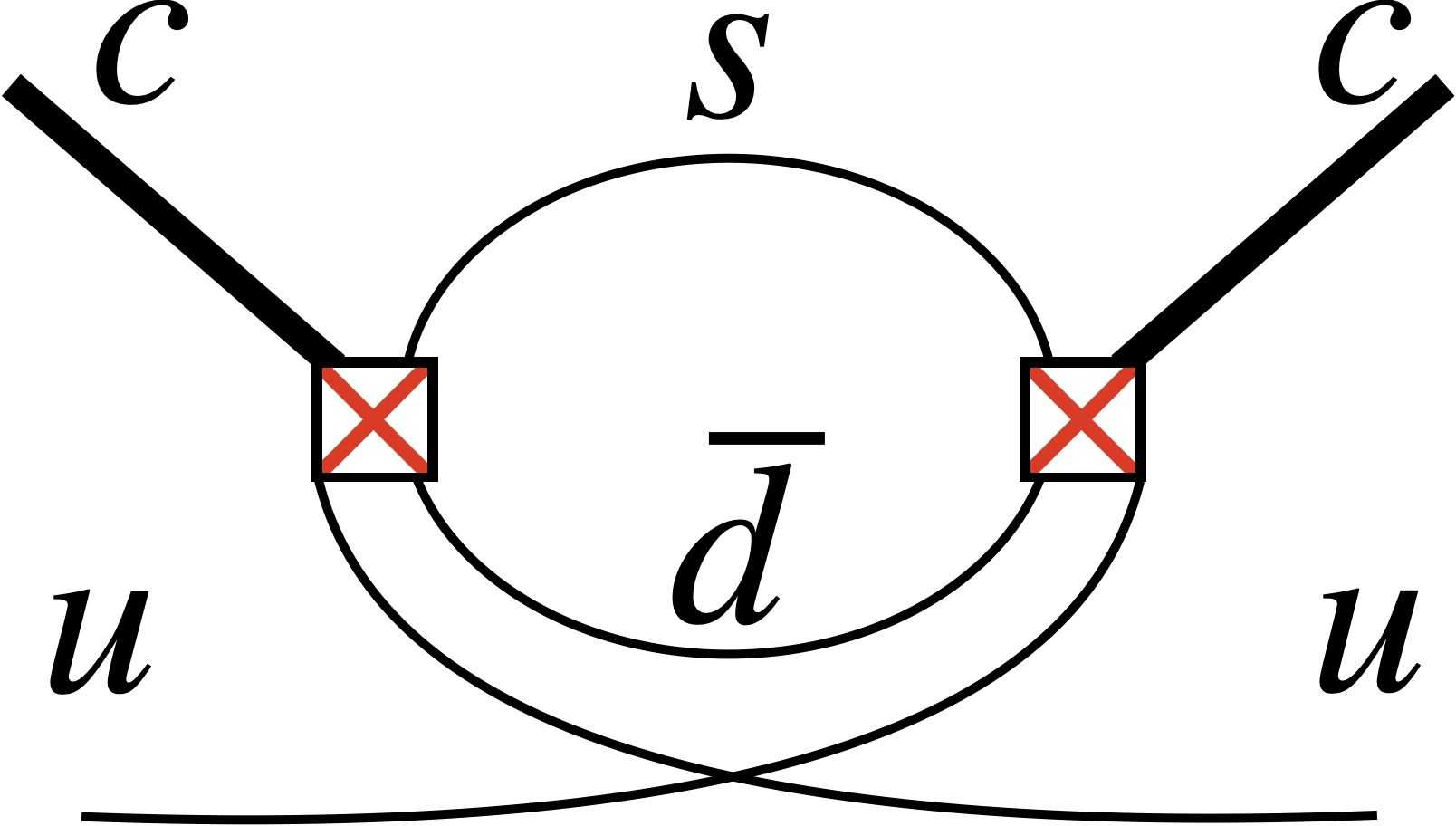}\\
				(b)
			\end{minipage}~~~~~
			\begin{minipage}{0.25\linewidth}
				\centering
				\includegraphics[width=\linewidth]{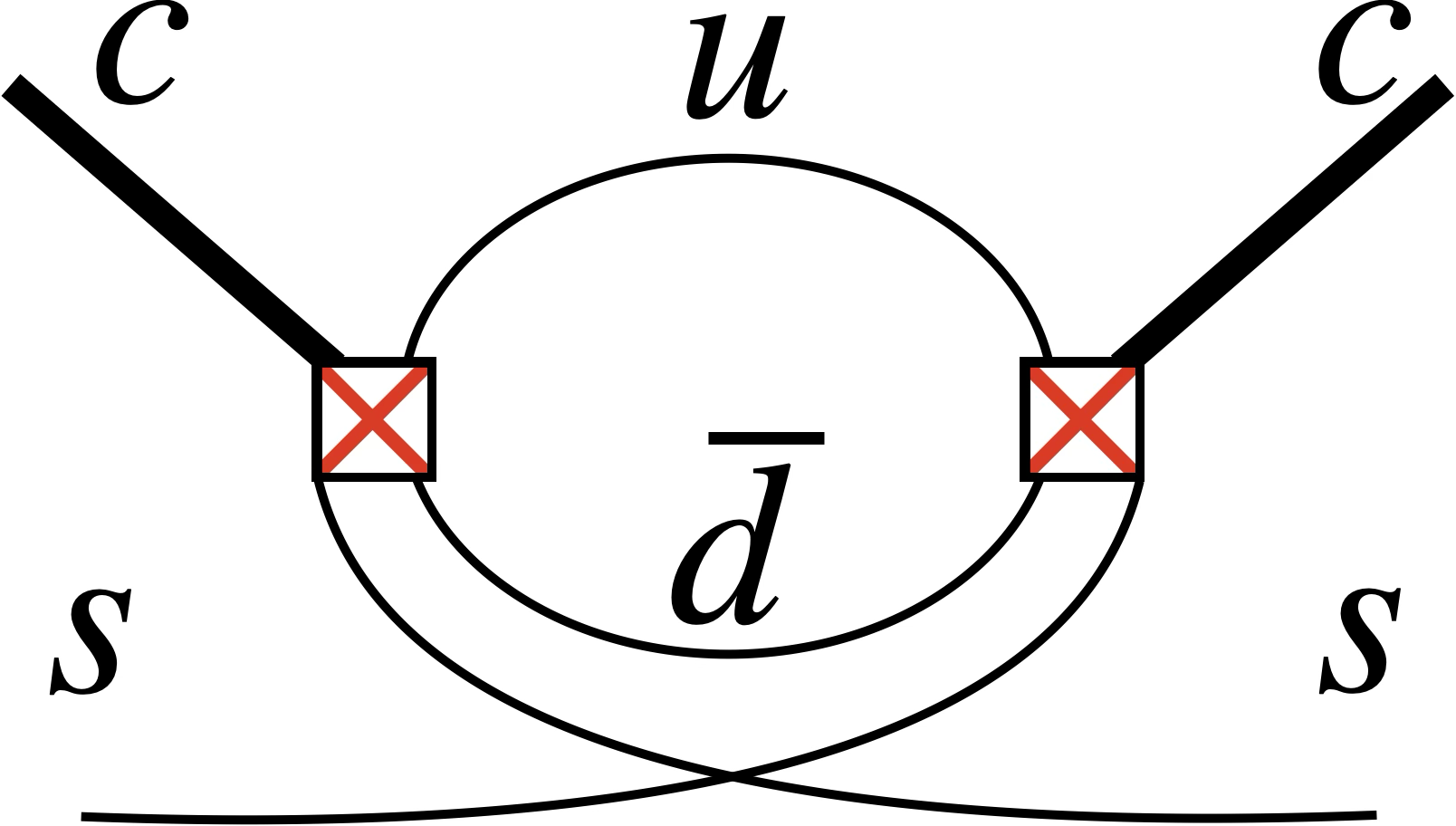}\\
				(c)
			\end{minipage} 
			\caption{
				The topological diagrams for the spectator effects: (a) \(W\)-exchange, (b) destructive (constructive) Pauli interference, and (c) constructive (destructive) Pauli interference at the dimension-6 (dimension-7) level.}
			\label{fig:spectator}
		\end{center}
	\end{figure}

	According to the optical theorem, the inclusive widths of heavy hadrons can be calculated from the matrix element of the transition operator ${\cal T}$, defined by 
	\begin{equation}
		\Gamma_{\mathrm{tot}} = 
		\frac{1}{2m _{\cal B} }
		\mathrm{Im} \left( i 
		\int \Big \langle T\left[ {\cal H}_{\rm eff} (x) {\cal H}_{\rm eff} (0 ) \right]  \Big \rangle d^4 x \right) 
		=
		\frac{1}{ 2 m _{\cal B} } \langle 
		{\cal B} | 
		{\cal T} | {\cal B}  \rangle
		\,,
	\end{equation}
	where \(T\) denotes the time-ordered product, \({\cal H}_{\rm eff}\) is the effective Hamiltonian obtained by integrating out the \(W\) boson, and   ${\cal B}$ represents either ${\cal B}_{cc}$ or ${\cal B}_{bb}$, namely the doubly charmed or doubly bottom baryons.
    The transition operator \({\cal T}\) is written symbolically as
	\begin{equation}\label{transition}
		\!\!\! \mathcal{T}=
		\frac{G_F^2 m_Q^5}{192\pi^3}\left[ 
		\left(\mathcal{C}_3 
		\overline{Q}Q 
		+\frac{\mathcal{C}_5}{m_Q^2} \mathcal{O}_5+\frac{\mathcal{C}_6}{m_Q^3} \mathcal{O}_6\right)_{\mathrm{2}}
		+16 \pi^2\left(\frac{ \tilde {\mathcal{C}}_6}{m_Q^3}  \tilde {\mathcal{O}}_6+\frac{\tilde{\mathcal{C}}_7}{m_Q^4} \tilde {\mathcal{O}}_7\right)_{\mathrm{4}}
		+\ldots 
		\right] , 
	\end{equation}
	where \({\cal C}\) denotes the Wilson coefficients and \({\cal O}\) the corresponding operators. The first parentheses contain two-quark operators, while the latter consist of four-quark operators, as indicated by the subscripts.
A phase-space factor of $16\pi^2$ is included for the latter, which can be further decomposed according to the light quark $q$; for instance 
\begin{equation}
		\tilde {\cal C} _6 \tilde {\cal O} _6  = \sum_{ {q}}
		\tilde 	{\cal C}_{q}  \tilde  {\cal O} _{q  }.
	\end{equation}
 In Cabibbo-favored charm decays \((c\to s \overline{d} u)\), the operators \(\tilde {\cal O}_{u,d,s}\) represent the topologies of destructive Pauli interference, \(W\)-exchange, and constructive Pauli interference, respectively, as depicted in Figure~\ref{fig:spectator}. In turn, each topology can be further categorized according to its Dirac structure:
	\begin{eqnarray}\label{WEHQE}
		\tilde 	{\cal C}_{q}  \tilde  {\cal O} _{q  } 
		&= &
		4		C_{q} ^{L}  ( \overline {Q}_\alpha  \gamma_\mu P_L 
		q _\alpha   ) 
		( \overline {q}_\beta   \gamma_\mu P_L 
		Q _\beta    ) 
		+4 \tilde C_{q} ^{L}   ( \overline {Q}_\alpha  \gamma_\mu  P_L 
		q _\beta    ) 
		( \overline {q}_\beta   \gamma_\mu  P_L 
		Q _\alpha     )   \nonumber \\ 
		&&+ 4  C_{q} ^{S}   ( \overline {Q}_\alpha  P_L 
		q _\alpha   ) 
		( \overline {q}_\beta  P_R 
		Q _\beta    ) 
		+4 \tilde C_{q} ^{S} ( \overline {Q}_\alpha  P_L 
		q _\beta    ) 
		( \overline {q}_\beta  P_R 
		Q _\alpha     )    \,. 
	\end{eqnarray}
	Here \(\alpha\) and \(\beta\) are colour indices. At dimension-6, all the light quarks are left-handed, since we can take \(m_q=0\). It is no longer the case once dimension-7 operators are considered.
	
	For transparency, we collect here the leading order~(LO) Wilson coefficients for the dimension-3, -5, and -6 operators. As an illustration, we consider only the case of
	\(c \to u \overline{d} s ,\)
	described by the effective Hamiltonian
	\begin{equation}
		\!\!\!\!\!{\cal H}_{\rm eff} 
		\!= \! \frac{4 G_F}{\sqrt{2}}
		V_{ud}V_{cs}^* 
		\left[
		c_1 ( \overline{s}  
		\gamma_\mu P_L
		c)
		( \overline{u}  
		\gamma^ \mu P_L
		d)
		+
		c_2 ( \overline{u}  
		\gamma_\mu  P_L 
		c)
		( \overline{s}  
		\gamma^ \mu P_L 
		d)
		\right]\,, 
	\end{equation}
	with \(V_{qq'}\) the CKM matrix elements, \(c_{1,2}\) the relevant Wilson coefficients, and summation over colour implied.  
With
$ 
		\xi=|V_{cs}|^2|V_{ud}|^2,$ $ x= ( m_s   / m_c) ^2,
$ 
	the leading two-quark Wilson coefficients are
	\begin{align}
		{\cal C}_3
		&=
		\xi \left(N_c c_1^2+N_c c_2^2+2c_1c_2\right)
		\Big[(1-x^2)(1-8x+x^2)-12x^2\ln x\Big],\\
		{\cal C}_5
		&=
		-\xi \left(N_c c_1^2+N_c c_2^2+2c_1c_2\right)(1-x)^4
		-8\xi\, c_1c_2(1-x)^3, 
	\end{align} with 
	${\cal O} _5
	= \overline{Q}
	\sigma_{\mu \nu} G^{\mu \nu}  Q/2 . 
	$ 
	For the four-quark operators, the coefficients are
	\begin{equation}\label{2.9}
		\!\!	C_d^L=12 \,\xi\,(1-x)^2\,c_1c_2,~~~~
		\tilde C_d^L = 6  \,\xi\,(1-x)^2(c_1^2+c_2^2), ~~~~
		C_d^{S}=0, ~~~~
		\tilde C_d^{S} =0, 
	\end{equation}  
	for \({\cal O}_d\), while
	\begin{align} 
		C_u^L&=-2 \,\xi\,(1-x)^2\left(1+\frac{x}{2}\right)(2c_1c_2+N_c c_2^2), &
		\tilde C_u^L&=-2 \,\xi\,(1-x)^2\left(1+\frac{x}{2}\right)c_1^2, \nonumber\\
		C_u^{S}&= 2 \,\xi\,(1-x)^2(1+2x)(2c_1c_2+N_c c_2^2), &
		\tilde C_u^{S}&= 2 \,\xi\,(1-x)^2(1+2x)c_1^2,
	\end{align} 
	for \({\cal O}_u\). Finally, the Wilson coefficients of \({\cal O}_s\) can be obtained from those of ${\cal O}_u$ by setting \(x \to 0\) and \(c_1 \leftrightarrow c_2\), as can be seen by comparing panels (b) and (c) in Figure~\ref{fig:spectator}:
	\begin{align}
		C_s^L&=-C_s^S=-2\,\xi\,(2c_1c_2+N_c c_1^2), &
		\tilde C_s^L&=-\tilde C_s^S=-2\,\xi\,c_2^2.
	\end{align}  
Strictly speaking the $x$ 
and $x^2$ 
appearing in the above formula shall be  treated as dimension-8 and dimension-10 corrections in a rigorous HQE treatment.  In this work  we   lump them into dimension-6 contributions for simplicity. 
We will see that the four-quark operators of the left-handed current have much larger matrix elements than the scalar ones. Based on the Wilson coefficients alone, 
Figure~\ref{fig:spectator}(a) is 
about six times larger than the others, making them the most important corrections at order \(m_Q^{-3}\). 
We will also see that the large factor of \(12\) in \(C_d^L\), together with the phase-space factor \(16\pi^2\), inverts the HQE hierarchy, so that the terms at order \(m_Q^{-3}\) exceed the leading contributions in charm decays. In the \(D\)-meson system, this contribution to the width is negative, and the inverted hierarchy is fatal, leading to a negative prediction for the \(D^+\) decay width at order \(m_c^{-3}\) and at LO in HQE~\cite{Cheng:1999fs}. Fortunately, in the baryon system the corresponding contribution is positive, making the situation less severe.
Expanding in \(\alpha_s\)  then leads to the NLO  corrections, but their explicit forms are too lengthy to present here.

For the dimension-7 operators $\tilde{\mathcal{O}}_7$ in HQE, we shall focus on the $1/m_Q$ corrections to the aforementioned spectator effects and neglect the operators with gluon fields. The dimension-7 four-quark operators encode the first \(1/m_Q\) corrections to the spectator effects from the dimension-6 sector. 
We shall follow \cite{Lenz:2013aua} to define the following dimension-7 four-quark operators:
	\begin{equation}\label{dim7}
		\begin{aligned}
			&P_1^q= m_q\,\overline{Q}(1-\gamma_5)q\,\overline{q}(1-\gamma_5)Q\,,\qquad
			P_2^q= m_q\,\overline{Q}(1+\gamma_5)q\,\overline{q}(1+\gamma_5)Q\,,\\
			&P_3^q= \frac{1}{m_Q}\,\overline{Q}\,\overleftarrow{D}_\rho \gamma_\mu(1-\gamma_5) D^\rho q\,\overline{q}\,\gamma^\mu(1-\gamma_5)Q\,,\\
			&P_4^q= \frac{1}{m_Q}\,\overline{Q}\,\overleftarrow{D}_\rho(1-\gamma_5)D^\rho q\,\overline{q}(1+\gamma_5)Q\,,
		\end{aligned}
	\end{equation}
	with \(\tilde P_i^q\) obtained by interchanging the colours of \(\overline q\) and \(Q\).
	At LO in \(\alpha_s\), we can discard \(A_\mu \) and replace 
    the full derivative $\partial_\mu $
    by \(\partial_t\), since the spatial derivative acting on the heavy quark is suppressed by \(1/m_Q\).
    For the explicit expressions of the spectator effects ${\cal T}_7^q$ with $(q=u,d,s)$ at the dimension-7 level, 
    the reader is referred to  Refs. \cite{Cheng:doubly,Dulibic:2023jeu} for details (see eq. (2.34) in \cite{Cheng:doubly} and Appendix B in \cite{Dulibic:2023jeu}).

	We stress that all the decay processes
	\(c\to s u \overline{d}, 	s u \overline{s}, d u \overline{d}, 	d u \overline{s}\)
	have been fully considered in this work for charm decays, and
	\(b \to c d \overline{c}, 	c d \overline{u}, u d \overline{c},u d \overline{u},
	c s \overline{c}, 	c s\overline{u}, u s \overline{c},u s  \overline{u}\)
	for bottom decays, with masses fully taken into account to NLO for dimension-3, -5, and -6 operators. For the dimension-7 operators, all decay processes will be considered at LO, since the NLO results are still missing in the literature. The leading Wilson coefficient \({\cal C}_3\) has been calculated to NLO for nonleptonic~(NL) decays~\cite{Bagan:1994zd,Bagan:1994qw,Krinner:2013cja}. For semileptonic~(SL) decays, the relevant expressions   can be found in Refs.~\cite{Mannel,Falk:1994,Moreno:2022goo}. On the other hand, the operators within \({\cal O}_5\) are closely related to the mass corrections of heavy hadrons from kinetic energy \((\mu_\pi^2)\) and gluonic interactions \((\mu_G^2 )\), with their explicit forms given later in eq.~\eqref{massoperator}. The corresponding Wilson coefficients for \(\mu_\pi^2 \) and \(\mu_G^2 \) have been calculated to NLO~\cite{Mannel:2015jka,Mannel:2023zei}. In recent years, the role of the Darwin operator associated with \({\cal O}_6\) has also been discussed. Its matrix elements can be related to those of the four-quark operators, and the corresponding Wilson coefficient has been studied in Refs.~\cite{Moreno:2022goo,NON_Bottom,Lenz:2020oce,King:2021xqp}. For the leading four-quark operator, \({\cal C}_6\) has been calculated to NLO, whereas only LO Wilson coefficients are currently available for \(\tilde{\cal O}_7\)~\cite{T6_NL,Lenz:2013aua,T71}.  
	
After collecting the Wilson coefficients, the remaining task is to calculate the matrix elements of each operator appearing in \({\cal T}\) within HQE. 
For the leading operator \(\overline{Q}Q\), one has
\begin{equation}
	\langle \overline{Q}Q \rangle _{\cal B} 
	= 
	2 - \frac{\mu_\pi^2 - \mu_G^2 }{2 m_Q^2 } + {\cal O} (m _Q ^{-3}) \,,
\end{equation}
with the shorthand notation
$\langle {\cal O}  \rangle _{\cal B} 
\equiv\langle {\cal B}| {\cal O} |{\cal B}  \rangle / (2 m_{\cal B}) \,.
$ 
This should be compared with the HQE mass formula
\begin{equation}\label{mugmupi}
	M_{{\cal B}} = 2 m _Q + \overline{\Lambda} 
	+\frac{\mu_\pi^2 }{2m_Q} - 
	C_G(m_Q, \mu) 
	\frac{\mu_G^2}{2m_Q} + O\bigl(1/m_Q^2\bigr)\,.
\end{equation} 
Here \(\overline{\Lambda}\) is the energy of the light quarks, and \(C_G(m_Q,\mu)\) is the Wilson coefficient reflecting the scale dependence of \(\mu_G^2\), connecting the scale \(\mu\) to \(m_Q\). 
	Due to the  reparametrization invariance~\cite{Abbott:1980hw}, $\mu_\pi^2$ is not renormalized in dimensional regularization.
In terms of the operators, 
the energy corrections are defined by 
\begin{eqnarray}\label{massoperator}
	\mu_\pi^2 &\equiv& \langle \overline{Q}_v (i\vec{D})^2 Q_v \rangle_{{\cal B}}=-\lambda_1+ O\bigl( 1/m_Q\bigr) \,, \nonumber\\
	\mu_G^2 &\equiv&  \frac{g_s}{2}\langle \overline{Q}_v \sigma_{\mu\nu}  G^{\mu\nu} Q_v \rangle_{{\cal B}} 
	=d_H {\lambda}_2 + O\bigl( 1/m_Q\bigr)\,, 
\end{eqnarray} 
with 
    \(
	D_\mu=\partial_\mu-ig_s A_\mu\,,
	\)
	and  \(A_\mu\)   the gluon field. 
The heavy-quark field \(Q_v\) is defined as
\begin{equation}\label{Eq3}
	Q_v(x)
	+ \frac{i\slash\hspace{-8pt}{D}_\perp}{2m_Q} Q_v (x) + O(1/m_Q^2)
	= e^{im_Q v\cdot  {x}} Q(x)\,,
\end{equation}
with $v$ the hadron velocity.   
After applying the HQE, 
there   remain several parameters that need to be calculated nonperturbatively. In this 
work, we will calculate all of the matrix elements 
within the BM, as detailed in the next section.

	\section{Baryonic matrix elements in the bag model }
	 
 In terms of creation operators, the baryon wave functions in the BM read 
	\begin{eqnarray}\label{struc}
		|{\cal B} _{QQ}, \uparrow\rangle &=& \int\frac{1}{2\sqrt{3} } \epsilon^{\alpha \beta \gamma} q _{a\alpha}^{\dagger} (\vec{x}_1) Q_{b\beta}^\dagger(\vec{x}_2) Q_{c\gamma}^\dagger (\vec{x}_3) \Psi_{qQQ}^{abc} (\vec{x}_1,\vec{x}_2,\vec{x}_3) [d^3  \vec{x}] | 0\rangle\,,
	\end{eqnarray}
	where $q$ denotes the light quark inside the doubly heavy baryons, Latin (Greek) letters denote spinor (colour) indices, and $\Psi$ describes the spinor--spatial distributions of the quarks. In the static bag limit, $\Psi$ is given by 
	\begin{eqnarray}\label{distri}
		\Psi_{ qQQ}^{abc } (
		\mathrm{static~bag};  
		\vec{x}_1,\vec{x}_2,\vec{x}_3)&=& 
		\frac{{\cal N}}{\sqrt{6}}\Big( 
		2 \phi^a_{q\uparrow}(\vec{x}_1)\phi^b_{Q\uparrow}(\vec{x}_2) \phi^c_{Q\downarrow}(\vec{x}_3)
		\\ 
		&&		- \phi^a_{q\downarrow}(\vec{x}_1)\phi^b_{Q\uparrow}(\vec{x}_2) \phi^c_{Q\uparrow}(\vec{x}_3)
		- \phi^a_{q\uparrow}(\vec{x}_1)\phi^b_{Q\downarrow}(\vec{x}_2) \phi^c_{Q\uparrow}(\vec{x}_3)
		\Big)\,,\non 
	\end{eqnarray}
	with ${\cal N}$ being a normalization constant. 
	The wave function is obtained by solving the boundary condition
\begin{equation}\label{tan}
	\tan({\bf p}_q R) = \frac{{\bf p}_q R}{ 1- m_q R - E^k_q R}\,,    
\end{equation}
	where the kinetic energy is $E^k _q = \sqrt{{\bf p}_q^2 + m_q^2}$. The   quark  wave function  is 
	\begin{equation}
		\phi_{q\updownarrow}(\vec{x})
		=
		\begin{pmatrix}
			u_q(\vec{x})\,\chi_{\updownarrow}\\
			i\,v_q(\vec{x})\,\hat{x}\cdot\vec{\sigma}\,\chi_{\updownarrow}
		\end{pmatrix},
	\end{equation} 
	with 
	\begin{equation}
		u_q(\vec{x})=
		\begin{cases}
			\omega_q^+\, j_0\!\left(|{\bf p}_q|\,|\vec{x}|\right), & \text{for } |\vec{x}|<R,\\
			0, & \text{otherwise},
		\end{cases}
		\quad
		v_q(\vec{x})=
		\begin{cases}
			\omega_q^-\, j_1\!\left(|{\bf p}_q|\,|\vec{x}|\right), & \text{for } |\vec{x}|<R,\\
			0, & \text{otherwise},
		\end{cases}
	\end{equation} 
	where   $\omega_q^\pm = \sqrt{E_q^k \pm m_q}$, $j_{0,1}$ are the spherical Bessel functions, $\chi_\uparrow = (1,0)^T $, $\chi_\downarrow = (0,1)^T$, and $R$ is the bag radius. 
The above BM equations also apply to the heavy quark after replacing $q$ with $Q$.

	The construction above clearly makes the baryon wave function a local object; the wave functions vanish for $|\vec{x}| >R$. According to the Heisenberg uncertainty principle, such a wave function cannot correspond to a momentum eigenstate. A straightforward improvement, without introducing any additional parameters, is to distribute the wave functions homogeneously:
	\begin{equation}
		\Psi_{ qQQ} ^{abc } (  
		\vec{x}_1,\vec{x}_2,\vec{x}_3)  
		= \int d^3 \vec x_{\Delta } 
		\Psi_{ qQQ} ^{abc } (
		\mathrm{static~bag};   
		\vec{x}_1 - \vec  x_{\Delta }  ,\vec{x}_2- \vec  x_{\Delta } ,\vec{x}_3 - \vec  x_{\Delta } ) \,. 
	\end{equation}
Employing this improved wave function in the calculation of the four-quark operators is essential for obtaining self-consistent results. For \(\mu_G^2\), however, we still use the original static BM approximation, since it is directly related to the hadron masses. We have checked that using the improved BM wave function instead leads to only minimal changes in the final numerical results of $\mu_G^2$.   
	
 In the BM, $R^{-1}$ serves as a characteristic hadronic energy scale, and the $R$-dependence of $\alpha_s$ is given by~\cite{Zhang:2021yul}
	\begin{equation}\label{alphas}
		\alpha_s = \frac{0.296}{\log\bigl(1+ (0.281R)^{-1}\bigr)}\,. 
	\end{equation}
	For the baryons $\Xi_{ cc}$ and $\Omega _{cc}$, one finds
	$\alpha_s = (0.50,\;0.51)$
	at
	$R = (4.42,\;4.49)\,\mathrm{GeV}^{-1}$, respectively.
	It is worth noting that $\alpha_s(1\,\text{GeV})=0.495$ in QCD~\cite{Herren:2017osy}, suggesting that the BM wave functions are applicable at the hadronic scale $\mu_H = 1$~GeV. In this work, we allow $\mu_H$ to vary within the range $0.8$ to $1.2$~GeV in order to account for hadronic uncertainties. 
	Since $\mu_\pi ^2$ corresponds to the kinetic-energy correction of the heavy quark to the baryon mass,  one has $\mu_\pi ^2=   2 {\bf p} _Q^2$ in the BM. 
    Taking the heavy quark limit in eq. (\ref{tan}) leads to $|{\bf p}_Q|=\pi/R$. The next-order correction has been worked out in \cite{Cheng:2023jpz} to be
  \begin{equation}
   |{\bf p}_Q|=\frac{\pi}{R}\left( 1- \frac{1}{2m_QR}+{\cal O}\left(\frac{1}{m^2_Q R^2}\right)\right).
   \end{equation}	
    On the other hand, $\mu_G$ has the physical meaning of the energy carried by the gluonic-field interference between quarks, and it is calculated by 
	\begin{equation}\label{14}
		\frac{\lambda_2}{2 m_Q} = 4 \pi \alpha_s \sum _{a} \int d^3 x \,\vec{B}_{Q\uparrow}^a (\vec{x}) \cdot \
        \left( \vec{B}_{Q\uparrow}^a (\vec{x})  + 
        \vec{B}_{q\uparrow}^a (\vec{x})\right)  \,,
	\end{equation} 
	with $d_H = 4 $.  The gluonic
	field inserted by $q$ is given by 
	\begin{equation}\label{magnetic}
		\vec{B}_{q }^a (\vec{x}) = \chi^{ \dagger}\Biggl(  \vec{\sigma}  \frac{\lambda^a}{4\pi}
		\left(
		2 M_q(r) + \frac{\mu_q(r)}{R^3} -\frac{\mu_q(r)}{r^3}  
		\right)
		+ \frac{3\lambda^a}{4\pi }\hat{r} ( \hat{r}\cdot \vec{\sigma}) \frac{\mu_q(r) }{ r^3 } \Biggr) \chi  \,,
	\end{equation}
	where the definitions of $M_q(r)$ and $\mu_q(r)$ can be found in Ref.~\cite{DeGrand:1975cf}, and $\lambda^a$ are the Gell-Mann matrices acting on the quark colour indices. 
The $\vec B_Q^a$ can be obtained by replacing $q$  with $Q$. 
    
	The 
	relevant 
	matrix elements of the dimension-6 operators are defined as  
	\begin{eqnarray}\label{4quark}
		&&L_{{\cal B} }^q \equiv 
		4 	
		\bigg \langle 
		(\overline {Q}_\alpha\gamma^\mu P_ L  q_\alpha )(\overline{q} _\beta \gamma_\mu P_ L  Q_\beta )\bigg \rangle_{{\cal B} } \,,\qquad \tilde{L}_{{\cal B}}^q  
		\equiv  4 \bigg \langle 
		(
		\overline {Q}_\beta \gamma^\mu P _L   q_\alpha )
		(\overline{q} _\alpha  \gamma_\mu P_L  Q_\beta )
		\bigg \rangle_{{\cal B}}
		\,,  \nonumber    \\
		&&S_{{\cal B}}^q \equiv \bigg \langle (\overline{Q}_\alpha q_\alpha )(\overline{q}_\beta Q_\beta )\bigg \rangle_{{\cal B}} \,,\qquad \tilde{S}_{{\cal B}}^q \equiv \bigg \langle (\overline{Q}_\alpha q_\beta  )(\overline{q}_\beta Q_\alpha )\bigg \rangle_{{\cal B}}\,,      \\
		&& P_{{\cal B}}^q \equiv \bigg \langle (\overline{Q}_\alpha \gamma_5 q_\alpha )(\overline{q}_\beta\gamma_5  Q_\beta )\bigg \rangle_{{\cal B}} \,,\qquad \tilde{P}_{{\cal B}}^q \equiv \bigg \langle (\overline{Q}_\alpha \gamma_5  q_\beta  )(\overline{q}_\beta \gamma_5  Q_\alpha )\bigg \rangle_{{\cal B}}\,.\nonumber
	\end{eqnarray}
	In the valence-quark approximation, valid around $\mu_H$, we have the relation $		I _{{\cal B}}^q   = - \tilde 		I _{{\cal B}}^q$ with $		I _{{\cal B}}^q = L	 _{{\cal B}}^q ,		S _{{\cal B}}^q ,	P _{{\cal B}}^q$. 
	The calculation of these matrix elements essentially involves computing the overlap of the two static bags separated by a distance $\vec x _\Delta$ and then integrating over $\vec x _\Delta$ to achieve translational invariance. A step-by-step derivation of these matrix elements is tedious and was documented in our previous work~\cite{Cheng:2023jpz,Cheng:2022jbr}. Here we give the final results 
		\begin{equation}\label{22}
			I _{{\cal B}}^q =
			\frac{{\cal N}^2}{ 2 M_{{\cal B}  }}
			\int d^3 \vec{x}_\Delta D_{Q}(\vec{x}_\Delta)\,  I  (\vec{x}_\Delta, {\cal B} )\,,
		\end{equation}
where $D_Q(\vec x _\Delta ) $ is the overlap coefficient of the spectator heavy quark, which does not participate in the weak interaction, and is given by
\begin{equation}
	D_Q \left(\vec{x}_{\Delta}\right)=  \int d^3 \vec{x} \,\phi_Q^{\dagger}\left(\vec{x}^+\right) \phi_Q\left(\vec{x}^-\right)
	=\int d^3 \vec{x}\left[u_Q^{+} u_Q^{-}+v_Q^{+} v_Q^{-}\left(\hat{x}^{+} \cdot \hat{x}^{-}\right)\right]\,,
\end{equation}
with $\vec{x}^\pm = \vec{x} \pm \vec{x}_\Delta /2$ and $|\hat{x}^\pm| = 1$.
We note that the appearance of $D_Q(\vec{x}_\Delta)$ indicates that the third, noninteracting quark enters as a weight function, which distinguishes this formalism from the static BM case.
The formalism reduces to the original static BM after dropping the integrals over $\vec{x}_\Delta$ and setting $\vec{x}_\Delta = 0$.
In that case, $D_Q(0)$ can be absorbed into ${\cal N}$, and the matrix elements become independent of the third noninteracting quark.

	The scalar matrix elements are 
	\begin{eqnarray}\label{in3}
		S (\vec{x}_\Delta,{{\cal B}} ) &=&
		\int d^3 \vec{x} \Big[ 
		{\cal C}_{\text{unflip}} 	\left( 
		u_Q^+ u_q^- -  v_Q^+v_q^-  
		\hat{x}^+ \cdot \hat{x}^-  
		\right)\left( 
		u_q^+ u_Q^- -  v_q^+v_Q^-  
		\hat{x}^+ \cdot \hat{x}^-   
		\right)  \nonumber\\
		&& \qquad
		-{\cal C}_{\text{flip}} \frac{(\vec{x}_\Delta \times \vec{x}  ) ^2}{(r^+r^-)^2}  v_Q^+v_q^- v_q^+v_Q^- \Big]\,, \nonumber\\
		P  (\vec{x}_\Delta,{{\cal B}} ) &=& - {\cal C}_{\text{flip}}  \int d^3 \vec{x}    \left(   u_Q^+v_q^- \hat{x}^- +  v_Q^+  u_q^-  \hat{x}^+ \right) \cdot 
		\left( u_q^+  v_Q^-  \hat{x}^- +  v_q^+  u_Q^-  \hat{x}^+ \right)\,. 
	\end{eqnarray}
	On the other hand, the left-handed current is 
	\begin{equation}\label{col1}
		L (\vec{x}_\Delta, {\cal B})  = \int d^3\vec{x} \sum_{k=1}^{4} \Upsilon^{{\cal B}}_k(\vec{x}_\Delta,\vec{x})\,,
	\end{equation}
	where 
		\begin{eqnarray}
		\Upsilon_1^{{\cal B}}(\vec{x}_\Delta,\vec{x}) &=&
		{\cal C}_{\text{unflip}} 	\left( 
		u_Q^+ u_q^- +  v_Q^+v_q^-  
		\hat{x}^+ \cdot \hat{x}^-  
		\right)\left( 
		u_q^+ u_Q^- +  v_q^+v_Q^-  
		\hat{x}^+ \cdot \hat{x}^-   
		\right) \nonumber\\
		&&-{\cal C}_{\text{flip}} \frac{(\vec{x}_\Delta \times \vec{x}  ) ^2}{(r^+r^-)^2}  v_Q^+v_q^- v_q^+v_Q^- \,, \nonumber\\
		\Upsilon_2^{{\cal B}} (\vec{x}_\Delta,\vec{x}) &=&-{\cal C}_{\text{flip}}   \left(   u_Q^+v_q^- \hat{x}^- -  v_Q^+  u_q^-  \hat{x}^+ \right) 
		\left( u_q^+  v_Q^-  \hat{x}^- -  v_q^+  u_Q^-  \hat{x}^+ \right) \,,\nonumber\\
		\Upsilon_3^{{\cal B}}  (\vec{x}_\Delta,\vec{x})&=&  -\frac{ {\cal C}_{\text{unflip}}  } { {\cal C}^{{\cal B}}_{\text{flip}}}\Gamma_2^{{\cal B}} (\vec{x}_\Delta,\vec{x})
		-2 {\cal C}_{\text{flip}}   \left( u_Q^+  v_q^-  \hat{x}^- +  v_Q^+  u_q^-  \hat{x}^+ \right) \cdot
		\left( u_q^+  v_Q^-  \hat{x}^- +  v_q^+  u_Q^-  \hat{x}^+ \right)  \,,\nonumber\\
		\Upsilon_4^{{\cal B}}(\vec{x}_\Delta,\vec{x}) &=&  -{\cal C}_{\text{flip}}   \Big[ 3 u_Q^+u_q^- u_q^+ u_Q^- +
			v_Q^+v_q^- v_q^+ v_Q^-  \left( 2 +(\hat{x}^+ \cdot\hat{x}^-)^2 \right) \label{col2} \\
		&&-(u_Q^+u_q^- v_q^+ v_Q^- + v_Q^+v_q^- u_q^+ u_Q^-  ) \hat{x}^+\cdot \hat{x}^-\Big]+	 {\cal C}_{\text{unflip}}   v_Q^+ v_q^- v_q^+ v_Q^-  \frac{   (\vec{x}_\Delta \times \vec{x}) ^2 }{(r^+r^-)^2} \,. \nonumber
	\end{eqnarray} 
	For the doubly heavy baryons, the spin overlap is 
	$({\cal C}_{\text{unflip}}, {\cal C}_{\text{flip}}) = (-1, 5/3)\,.     $
	Taking $m_Q\to \infty$, we have $v_Q \to 0$ and 
	\begin{equation}
		L_{{\cal B}} ^q = 6 S_{{\cal B}}^q + \frac{2}{5} P_{{\cal B}}^q\,,
	\end{equation}
Taking the limit \(m_q \to \infty\), this relation reduces to the nonrelativistic quark-model result \(L_{\cal B}^q = 6 S_{\cal B}^q\)~\cite{Cheng:2022jbr,Cheng:doubly,Dulibic:2023jeu}.

The numerical results for the matrix elements are collected in Table~\ref{tab:4quark_DHB}, together with the corresponding bag radii listed in the first column. Both are obtained using the BM parameters \((m_{u,d},m_s,m_c,m_b)=(0,0.279,1.641,5.093)\,\mathrm{GeV}\) from Ref.~\cite{Zhang:2021yul}. The bag radii of \(\Xi_{bb}\) and \(\Omega_{bb}\) are about \(20\%\) smaller than those of \(\Xi_{cc}\) and \(\Omega_{cc}\). Since dimensional analysis implies \(\mu_\pi^2,\mu_G^2\sim R^{-2}\) and \(I_{\cal B}^q\sim R^{-3}\), the dimension-6 matrix elements of the doubly bottom baryons become nearly twice as large as those of the doubly charmed baryons.
This feature is potentially problematic, because heavy-quark flavor symmetry suggests that
\begin{equation}
	I_{\Xi_{bb}}^q =
	I_{\Xi_{cc}}^q
	+ O(1/m_c) \,.
\end{equation}
Indeed, in heavy quark effective theory {(HQET)} the leading-order Lagrangian is
\begin{equation}
{\cal L}=\overline{h}_v\, i v\!\cdot\! D\, h_v + O(1/m_Q),
\end{equation}
with \(h_v=(c_v,b_v)^T\). Since this Lagrangian is invariant under heavy-flavor \(SU(2)_Q\) transformations, one finds, for example,
\begin{equation}
\langle (\overline{c}_v q)(\overline{q} c_v)\rangle_{{\cal B}_{cc}}
=
\langle (\overline{b}_v q)(\overline{q} b_v)\rangle_{{\cal B}_{bb}} .
\end{equation}
This simply reflects the fact that, in the heavy-quark limit, charm and bottom quarks act as identical static color sources and hence interact with the light quarks in the same way. Similar relations apply to \(\mu_\pi^2\) and \(\mu_G^2\).
	The sizable departure from this expectation motivates an estimate of the systematic uncertainty associated with the bag radii extracted directly from the mass spectra. For this reason, we consider two scenarios in the present work. In the first, we use the radii obtained from the mass spectra, as listed in the first row of Table~\ref{tab:4quark_DHB}. In the second, we adopt a common value \(R=4.6~\mathrm{GeV}^{-1}\) for all doubly heavy baryons. This choice is motivated by the \(\Lambda_b\) baryon, for which \(R=4.6~\mathrm{GeV}^{-1}\) is a commonly used bag radius in phenomenological studies of heavy-baryon decays. 

		The spectrum-derived radii used as central values represent the genuine bag-model predictions. The common-radius scenario, being closer to the heavy-quark-flavor-symmetry limit, is adopted as a physically motivated systematic variation to estimate the size of heavy-flavor-symmetry breaking. As shown in Table~\ref{tab:4quark_DHB}, using the bag radii extracted from the mass spectra, the ratios
$
\langle O\rangle_{{\cal B}_{cc}}/\langle O\rangle_{{\cal B}_{bb}}
$
for the dominant four-quark matrix elements are typically around $60\%$; for example,
$
L_{\Xi_{cc}}^q/L_{\Xi_{bb}}^q \simeq 0.63.
$
For comparison, the analogous ratio in the heavy-meson case is
$
f_D^2 m_D/(f_B^2 m_B)\simeq 0.44,
$
where unity would be expected in the heavy-quark-symmetry limit. Thus, the heavy-flavor-symmetry breaking induced by the different bag radii in doubly heavy baryons is sizable but comparable to, or smaller than, the familiar breaking between the $D$- and $B$-meson systems.
{This comparison also shows that the near heavy-flavor stability of \(L_{{\cal B}_Q}^q\) found for singly heavy baryons in the BM does not persist quantitatively for doubly heavy baryons; the resulting \(\sim 40\%\) breaking should be kept in mind in the HQET-counting arguments used below.}

	\begin{table}[t]
		\centering
		\caption{Four-quark matrix elements and the kinetic and chromomagnetic parameters of doubly heavy baryons at the hadronic scale $\mu_H$. The entries for $L_q^{\cal B}$, $S_q^{\cal B}$, and $P_q^{\cal B}$ are given in units of $10^{-3}\,{\rm GeV}^3$, while $\mu_\pi^2$ and $\mu_G^2$ are given in units of $10^{-1}\,{\rm GeV}^2$. For each baryon, the first subcolumn gives the spectrum-derived BM result, and the second gives the result obtained with the common radius $R=4.6~{\rm GeV}^{-1}$. We use the former as the central input and the latter only as a heavy-flavor-symmetry motivated systematic comparison.}
		\label{tab:4quark_DHB}
        \begin{center}
		\begin{tabular}{c|rr|rr|rr|rr}
			\hline
			\hline
			& \multicolumn{2}{c|}{$\Xi_{cc}$}
			& \multicolumn{2}{c|}{$\Omega_{cc}$}
			& \multicolumn{2}{c|}{$\Xi_{bb}$}
			& \multicolumn{2}{c}{$\Omega_{bb}$} \\
			$R$
			& $4.42$ & $4.60$
			& $4.49$ & $4.60$
			& $3.71$ & $4.60$
			& $3.83$ & $4.60$ \\
			\hline 
			$\mu_\pi^2$
			& $8.82$ & $8.19$
			& $8.57$ & $8.19$
			& $13.60$ & $8.94$
			& $12.78$ & $8.94$ \\
			$\mu_G^2$
			& $2.44$ & $2.34$
			& $2.02$ & $1.96$
			& $3.39$ & $2.61$
			& $2.83$ & $2.20$ \\
			\nonumber 
			$L^q_{\cal B}$
			& $-37.45$ & $-33.12$
			& $-41.02$ & $-38.22$
			& $-59.50$ & $-30.92$
			& $-63.15$ & $-37.20$ \\
			$S^q_{\cal B}$
			& $-5.14$ & $-4.57$
			& $-5.95$ & $-5.56$
			& $-9.07$ & $-4.78$
			& $-9.87$ & $-5.89$ \\
			$P^q_{\cal B}$
			& $-2.06$ & $-1.78$
			& $-1.59$ & $-1.45$
			& $-1.97$ & $-0.93$
			& $-1.44$ & $-0.72$ \\
			\hline
			\hline
		\end{tabular}
        \end{center}
	\end{table} 
	
As for the dimension-7 four-quark operators given in eq. (\ref{dim7}), their matrix elements in the BM can be expressed in terms of that of dimensions-6 ones	:	
	\begin{equation}
		\langle P_3^q\rangle_{{\cal B}} =  E_q\,L_{{\cal B}}^q,
        \qquad
		\langle P_4^q\rangle_{{\cal B}} =  E_q\left(S_{{\cal B}}^q-P_{{\cal B}}^q\right), 
	\end{equation}
up to $O (\alpha_s)$, where we have applied the approximation \cite{Cheng:2023jpz}
\begin{eqnarray}\label{approx}
		&&\frac{1}{m_Q}\left \langle \left( \overline{Q} \overleftarrow{D}_\rho \Gamma D^\rho q\right)  \overline{q} \Gamma  Q\right \rangle \approx E_q \left \langle  \overline{Q}   \Gamma   q  \overline{q} \Gamma  Q\right \rangle \,,
	\end{eqnarray}
with $E_q$ being the energy of the bag quark. 
	We will follow \cite{Cheng:2023jpz} to use $(E_{u,d} , E_s)  = 
	( 0.32, 0.50)
	$~GeV.

	The Darwin operator provides the third two-quark contribution in HQE and enters the inclusive width at order \(1/m_Q^3\) through
	\begin{equation}
		\Gamma({\cal B} )\supset 
		\frac{G_F^2m_Q^2}{192\pi^3}\,
		{\cal C}_\rho \rho _D^3 \,,~~~ \rho _D^3 \equiv 
		\left\langle 
		\overline{Q}_v (iD_\mu)(iv\!\cdot\!D)(iD^\mu)Q_v
		 \right\rangle_{\cal B}\,.
	\end{equation} 
	In the  analysis, \(\rho_D^3\) is  related to the dimension-6 four-quark matrix elements through~\cite{Bigi:1993ex}
	\begin{equation}
		\rho_D^3
		=
		-4\pi\alpha_s  \frac{1}{24}
		\left(
		4L_{{\cal B}}^q-\tilde L_{{\cal B}}^q
		-6S_{{\cal B}}^q+2\tilde S_{{\cal B}}^q
		+6P_{{\cal B}}^q-2\tilde P_{{\cal B}}^q
		\right),
	\end{equation}
	where the equation of motion has been applied.  
Since $\rho_D^3$ is proportional to $\alpha_s$, we take it to be zero at LO.

	The matrix elements entering the heavy-baryon lifetimes are evaluated in the BM at a hadronic scale \(\mu_H\sim 1~\mathrm{GeV}\), and then evolved to the heavy-quark scale \(\mu_Q\sim m_Q\). Among the two-quark operators, \(\mu_\pi^2\) is protected by reparametrization invariance and does not run in dimensional regularization. By contrast, \(\mu_G^2\) is scale dependent, but its running is compensated by the Wilson coefficient \(C_G(m_Q,\mu)\), so that the product \(C_G(m_Q,\mu)\mu_G^2(\mu)\) is renormalization-scheme independent. Accordingly, one uses~\cite{Grozin:2007fh}
	\begin{equation}\label{3.20}
C_G(m_Q,\mu_H)\,\mu_G^2(\mu_H)=C_G(m_Q,m_Q)\,\mu_G^2(m_Q)\,,
	\end{equation}
	with \(C_G(m_c,m_c)=1.6506\) and \(C_G(m_b,m_b)=1.2664\).  
	In the BM, \(C_G(m_Q,\mu_H) =1\) is used to match the mass spectra. 
	
	For the dimension-6 four-quark matrix elements, the BM results are first obtained at \(\mu_H\), where the valence-quark approximation gives $		I _{{\cal B}}^q(\mu_H)   = - \tilde 		I _{{\cal B}}^q(\mu_H) $. They are then evolved to \(\mu_Q\). In HQET, the leading-logarithmic evolution reads~\cite{Neubert:1996we}
	\begin{equation}
		I_{{\cal B}}^q(\mu_Q)=\kappa\, I_{{\cal B}}^q(\mu_H)-\frac{1}{3}(\kappa-1)\,\tilde I_{{\cal B}}^q(\mu_H),\qquad
		\tilde I_{{\cal B}}^q(\mu_Q)=\tilde I_{{\cal B}}^q(\mu_H),
	\end{equation}
	with \(\kappa=\sqrt{\alpha_s(\mu_H)/\alpha_s(\mu_Q)}\). Numerically, this gives \(I_{{\cal B} }^q(\mu_b)\approx(1.56\pm0.19)I_{{\cal B} }^q(\mu_H)\) and \(I_{{\cal B} }^q(\mu_c)\approx(1.26\pm0.15)I_{{\cal B} }^q(\mu_H)\). 
	In the full-QCD \(\overline{\rm MS}\) scheme, the evolution of the four-quark matrix elements is 
	\begin{equation}
 I_{{\cal B}}^q(\mu_Q)
        -\tilde I_{{\cal B}}^q(\mu_Q) 
		=
		U_-(\mu_H,\mu_Q)\,
        \Big(I_{{\cal B}}^q(\mu_H)
        -\tilde I_{{\cal B}}^q(\mu_H)\Big)
        \,.
	\end{equation}
	Meanwhile, we  have
	\(
	\tilde I_{{\cal B}}^q(\mu_Q)+I_{{\cal B}}^q(\mu_Q)
    = U _+ (\mu_H , \mu _Q ) 
(\tilde I_{{\cal B}}^q(\mu_H)+I_{{\cal B}}^q(\mu_H)) = 
    0
	\).  	
In the naive dimensional regularization (NDR) scheme at NLO, the results are~\cite{Buchalla:1995vs}\footnote{
At NLO, the renormalization-group evolution of the operators induces a new set of penguin operators, namely eye-diagram contributions~\cite{King:2021jsq,Black:2026rbz} involving right-handed light quarks. 
Their contributions cancel in the decay-width differences, and we neglect this type of contraction throughout this work. 
} 
	\begin{eqnarray}\label{51}
		U_-^{\rm NLO}(\mu_H,\mu_c)
		&=&
		\left[
		\frac{\alpha_s(\mu_H)}{\alpha_s(\mu_c)}
		\right]^{\frac{4}{9}}
		\left[
		1-\frac{181}{81}\,
		\frac{\alpha_s(\mu_H)-\alpha_s(\mu_c)}{4\pi}
		\right]
		\,, \\
		U_-^{\rm NLO}(\mu_H,\mu_b)
		&=&
		U_-^{\rm NLO}(\mu_H,\mu_c) 
		\left[
		\frac{\alpha_s(\mu_c)}{\alpha_s(\mu_b)}
		\right]^{\frac{12}{25}}
		\left[
		1-\frac{3569}{1875}\,
		\frac{\alpha_s(\mu_c)-\alpha_s(\mu_b)}{4\pi}
		\right]
		\,. \non
	\end{eqnarray} 
	At LO, one sets
	the term proportional to 
	$\alpha_s(\mu_c) - \alpha_s(\mu_{b,H})$
	to zero. 
	Taking $ \mu_H =( 1.0 \pm 0.2) \text{ GeV}$,  
	we have 
	$	U_-^{\rm NLO}(\mu_H,\mu_c)  = 1.16\pm 0.09 $ 
	and 
	$	U_-^{\rm NLO}(\mu_H,\mu_b)  = 1.42\pm 0.11. $ 
	
\section{Numerical results}

We adopt the same parameters as in our previous work and briefly summarize them here for completeness. The Wilson coefficients of the effective Hamiltonian ${\cal H}_{\rm eff}$ are taken in the NDR scheme. For charm-quark decays at $\mu_c = 1.5~\text{GeV}$, we use the NLO values~\cite{Buchalla:1995vs}
\begin{equation}
(c_1, c_2) = (1.188, -0.378)\,,
\end{equation}
while for bottom-quark decays at $\mu_b = 4.4~\text{GeV}$, we take
\begin{equation}
(c_1, c_2, c_3, c_4, c_5, c_6) = (1.078, -0.184, 0.013, -0.030, 0.009, -0.038)\,.
\end{equation}
Since the penguin coefficients are proportional to $\alpha_s$, we set them to zero at LO, namely $c_{3,4,5,6}=0$, while the leading coefficients are taken to be $(c_1, c_2) = (1.139, -0.301)$ at $\mu_b$ and $(1.298, -0.565)$ at $\mu_c$.  In much of the literature, the uncertainties associated with the energy scale of the matrix elements are estimated by varying the Wilson-coefficient scale $\mu_Q$. In this work, we take the opposite approach: we vary the renormalization scale $\mu_H$ of the hadronic matrix elements and evolve them to a fixed $\mu_Q$, as shown in Eqs.~\eqref{3.20} and \eqref{51}. The two prescriptions are equivalent.

The pole masses used in eq.~\eqref{transition} are
\begin{equation}
(m_c, m_b)_{\mathrm{pole}} = (1.59 \pm 0.09, 4.70 \pm 0.10)~\text{GeV},
\end{equation}
where the uncertainties correspond to the difference between the one‑loop and two‑loop conversion from the $\overline{\mathrm{MS}}$ masses $\overline{m}_c(\overline{m}_c) = 1.27~\text{GeV}$.  
Other heavy-quark mass schemes differ from the pole mass only by perturbative corrections in \(\alpha_s\), and the ranges above are broad enough to cover the values obtained in commonly used schemes.   We use the same heavy‑quark masses for both LO and NLO calculations because, within HQE, the heavy quark is nearly static inside the hadron and the pole mass has historically been used as input.  However, free heavy quarks do not exist and the pole mass suffers from renormalon ambiguities~\cite{Beneke:1998ui}; the pole masses should therefore be regarded as phenomenological inputs rather than fundamental definitions.  For the light‑quark masses in loops we use $\overline{\mathrm{MS}}$ values,
\begin{equation}
\bigl(\overline{m}_c(\overline{m}_c), \overline{m}_s(2~\text{GeV})\bigr) = \bigl(1.27, 0.09\bigr)~\text{GeV},
\end{equation}
consistent with taking the Wilson coefficients in the $\overline{\mathrm{MS}}$ scheme.

The numerical results are collected in 
Table~\ref{tab:LO_NLO}. HQE converges much more reliably for ${\cal B}_{bb}$  than for ${\cal B}_{cc}$.  In the bottom sector the dimension-3 contributions dominate and higher‑dimensional terms behave as genuine corrections.  In the charm sector the convergence is visibly slower: although HQE appears acceptable for $\Xi_{cc}^{++}$, it is much less well behaved for $\Xi_{cc}^+$ and $\Omega_{cc}^+$, where the dimension-6 spectator effects are very large and can even exceed the leading contribution. {For \(\Xi_{cc}^{+}\), this large term corresponds to the \(W\)-exchange topology shown in Figure~\ref{fig:spectator}(a). The comparison with the familiar \(D^+\) problem is topology dependent: the same \(q=d\) spectator-operator channel appears as destructive Pauli interference in \(D^+\), where the truncated HQE width can become negative, but as constructive \(W\)-exchange in \(\Xi_{cc}^{+}\).} Nevertheless, the expansion is not entirely pathological, since the dimension-7 terms remain smaller than the dimension-6 ones.  Contributions from even higher-dimensional operators, such as dimension-9 six-quark operators, are further suppressed by $\alpha_s$, so an inverted hierarchy is not expected.

The dominant dimension-6 contribution arises from the $W$-exchange topology, corresponding to processes such as $cd \to cd$ in the charm sector and $bu \to bu$ in the bottom sector. This mechanism gives rise to the especially large spectator effects in $\Xi_{cc}^+$ and $\Xi_{bb}^0$. In both sectors, the sizes of the corrections are much larger than one would naively expect from simple power counting based on $16\pi^2 (\Lambda_{\rm QCD}/m_Q)^3$, indicating that the structure of the hadronic matrix elements plays a far more important role than naive dimensional analysis alone would suggest. 
In addition to the large Wilson coefficients  discussed in Sec.~\ref{sec2},
this large enhancement can be traced to the spin-flavor structure of doubly heavy baryons: in the quark model, ${\cal B}_{cc}$ and ${\cal B}_{bb}$ contain two identical valence heavy quarks, in contrast to singly heavy baryons such as $\Lambda_Q$, whose three valence quarks are all different.
In particular, ${\cal C}_{\mathrm{flip}}$, which denotes the overlap of the spin-flavor wave functions in eq.~\eqref{col2}, is more than three times larger for doubly heavy baryons than for the $\Lambda_Q$ case. For the same reason, HQE for $\Omega_c^0$ appears to be less well controlled~\cite{Cheng:2023jpz}, where the dimension-6 contribution is also larger than the leading one.

\begin{landscape}
	\vspace{-2cm}
	\begin{table}
		\caption{ 
The decay widths and lifetimes are given in units of $10^{-12}~\text{GeV}$ and $10^{-13}~\mathrm{s}$ for ${\cal B}_{cc}$, and in units of $10^{-13}~\text{GeV}$ and $10^{-12}~\mathrm{s}$ for ${\cal B}_{bb}$, respectively. 
Here, $\Gamma_{3}$ stands for the contribution from the leading operator $\overline{Q}Q$, $\Gamma_\rho$ from ${\cal O}_6$, and $\Gamma_{6,7}$ from $\tilde{\cal O}_{6,7}$ in eq.~\eqref{transition}. The superscripts NL and SL denote the contributions from nonleptonic and semileptonic decays, respectively. ${\cal BF}_e^{\mathrm{SL}}$ stands for the inclusive branching fraction of ${\cal B}\to X e$. 
   The central values are evaluated in the full-QCD scheme with the spectrum-derived BM matrix elements \(I_{\cal B}^q\). The parentheses are ordered as follows: the first and second parentheses are the symmetric uncertainties from \(m_Q\) and \(\mu_H\), respectively; the third parenthesis is the signed shift obtained by replacing the spectrum-derived radii with the common value \(R=4.6~\mathrm{GeV}^{-1}\); and the fourth parenthesis is the signed shift obtained by using the HQET scheme. Thus, the last two parentheses are diagnostic systematic shifts, not independent symmetric errors. For example, \(25.63(2.45)(1.14)(-1.78)(-1.18)\) means \(25.63\pm2.45\pm1.14\), plus a shift \(-1.78\) from the common-radius choice and a shift \(-1.18\) from the HQET scheme. The NLO values of $\Gamma_7$ are taken to be the same as the LO ones.}
		\label{tab:LO_NLO}
		\vspace{6pt}
		\centering
		\setlength{\tabcolsep}{5pt}
		\begin{tabular}{lc|ccccccccc}
			\hline\hline
			${\cal B}_{QQ}$ & & $\Gamma_3^{\rm NL}$ & $\Gamma_3^{\rm SL}$ & $\Gamma_{\rho}$ & $\Gamma_6^{\rm NL}$ & $\Gamma_6^{\rm SL}$ & $\Gamma_7^{\rm NL}$ & $\Gamma_7^{\rm SL}$ & ${\cal BF}^{\rm SL}_e(\%)$ & $\tau$ \\
			\hline
			\multirow{2}{*}{$\Xi_{cc}^{++}$}
			& LO  & $1.77$ & $0.85$ &      & $-1.61$ & $0.00$ & $0.66$ & $-0.00$ & $25.63(2.45)(1.14)(-1.78)(-1.18)$ & $3.96(1.93)(17)(-33)(-18)$ \\
			& NLO & $2.04$ & $0.64$ & $0.10$ & $-0.97$ & $0.00$ & $0.66$ & $-0.00$ & $13.44(9)(13)(-22)(+62)$ & $2.67(94)(2)(-7)(+12)$ \\
			\hline
			\multirow{2}{*}{$\Xi_{cc}^{+}$}
			& LO  & $1.77$ & $0.85$ &      & $8.04$ & $0.07$ & $2.92$ & $-0.03$ & $3.27(49)(22)(+36)(+12)$ & $0.48(7)(4)(+5)(+2)$ \\
			& NLO & $2.04$ & $0.64$ & $0.10$ & $8.22$ & $0.03$ & $2.92$ & $-0.03$ & $2.40(36)(16)(+26)(0)$ & $0.47(7)(4)(+5)(0)$ \\
			\hline
			\multirow{2}{*}{$\Omega_{cc}^{+}$}
			& LO  & $1.76$ & $0.85$ &      & $2.88$ & $1.50$ & $-1.34$ & $-1.02$ & $14.41(46)(9)(+9)(+13)$ & $1.42(41)(5)(+4)(-3)$ \\
			& NLO & $2.04$ & $0.64$ & $0.10$ & $2.49$ & $0.76$ & $-1.34$ & $-1.02$ & $5.48(1.71)(39)(+37)(+88)$ & $1.79(62)(4)(+2)(+15)$ \\
			\hline
			\hline
			\multirow{2}{*}{$\Xi_{bb}^{0}$}
			& LO  & $4.68$ & $2.29$ &      & $1.02$ & $0.00$ & $0.16$ & $0.00$ & $12.74(4)(15)(+93)(+37)$ & $0.81(11)(1)(+5)(+2)$ \\
			& NLO & $5.41$ & $2.01$ & $-0.05$ & $1.23$ & $0.00$ & $0.16$ & $0.00$ & $10.39(3)(13)(+81)(+19)$ & $0.75(11)(1)(+6)(+1)$ \\
			\hline
			\multirow{2}{*}{$\Xi_{bb}^{-}$}
			& LO  & $4.68$ & $2.29$ &      & $-0.28$ & $0.00$ & $0.05$ & $0.00$ & $15.42(24)(4)(-29)(-11)$ & $0.98(16)(0)(-3)(-1)$ \\
			& NLO & $5.41$ & $2.01$ & $-0.05$ & $-0.24$ & $0.00$ & $0.05$ & $0.00$ & $12.69(20)(4)(-23)(-4)$ & $0.92(15)(1)(-2)(-1)$ \\
			\hline
			\multirow{2}{*}{$\Omega_{bb}^-$}
			& LO  & $4.69$ & $2.29$ &      & $-0.28$ & $0.00$ & $-0.08$ & $0.00$ & $15.69(28)(7)(-38)(-6)$ & $0.99(17)(0)(-2)(0)$ \\
			& NLO & $5.42$ & $2.02$ & $-0.05$ & $-0.23$ & $0.00$ & $-0.08$ & $0.00$ & $12.89(23)(6)(-29)(+1)$ & $0.93(15)(0)(-2)(0)$ \\
			\hline
			\hline
		\end{tabular}
	\end{table}
\end{landscape}
\clearpage
\newpage

Combining the positive and negative variations separately in quadrature gives the NLO lifetime estimates
\begin{equation}
(\tau_{\Xi_{cc}^{++}},\tau_{\Xi_{cc}^{+}},\tau_{\Omega_{cc}^{+}})
=
\left(2.67^{+0.95}_{-0.94},\,0.47^{+0.09}_{-0.08},\,1.79^{+0.64}_{-0.62}\right)
\times 10^{-13}\,{\rm s},
\end{equation}
\begin{equation}
(\tau_{\Xi_{bb}^{0}},\tau_{\Xi_{bb}^{-}},\tau_{\Omega_{bb}^{-}})
=
\left(0.75^{+0.13}_{-0.11},\,0.92^{+0.15}_{-0.15},\,0.93^{+0.15}_{-0.15}\right)
\times 10^{-12}\,{\rm s}.
\end{equation}
These compact uncertainties summarize the detailed error decomposition displayed in Table~\ref{tab:LO_NLO}. The same directional quadrature prescription is used for the semileptonic widths in Table~\ref{tab:semi} and for the asymmetries below.
{The quoted uncertainties combine the parameter variations and diagnostic radius/scheme shifts specified above; in this combination, the signed diagnostic shifts are treated as one-sided error proxies in their corresponding directions. An additional independent uncertainty from the truncation of the HQE is not included. This convention is particularly relevant for \(\Xi_{cc}^{+}\), for which the dimension-6 contribution exceeds the nominally leading term.}

To quantify the spectator effects, we define  the decay width  asymmetries
\begin{equation}\label{Rcc}
	{\cal R}_{cc}
	=
	\frac{
		\tau_{\Xi_{cc}^{+}}^{-1}
		-\tau_{\Xi_{cc}^{++}}^{-1}
	}{
		\tau_{\Xi_{cc}^{+}}^{-1}
		+\tau_{\Xi_{cc}^{++}}^{-1}
	}\,,
	\qquad
	{\cal R}_{cc}^{\Omega}
	=
	\frac{
		\tau_{\Omega_{cc}^{+}}^{-1}
		-\tau_{\Xi_{cc}^{++}}^{-1}
	}{
		\tau_{\Omega_{cc}^{+}}^{-1}
		+\tau_{\Xi_{cc}^{++}}^{-1}
	}\,,
\end{equation}
\begin{equation}
	{\cal R}_{bb}
	=
	\frac{
		\tau_{\Xi_{bb}^{0}}^{-1}
		-\tau_{\Xi_{bb}^{-}}^{-1}
	}{
		\tau_{\Xi_{bb}^{0}}^{-1}
		+\tau_{\Xi_{bb}^{-}}^{-1}
	}\,,
	\qquad
	{\cal R}_{bb}^{\Omega}
	=
	\frac{
		\tau_{\Xi_{bb}^{0}}^{-1}
		-\tau_{\Omega_{bb}^{-}}^{-1}
	}{
		\tau_{\Xi_{bb}^{0}}^{-1}
		+\tau_{\Omega_{bb}^{-}}^{-1}
	}\,,
\end{equation}
bounded between $-1$ to $1$. 
In particular, they vanish  in the absence of dimension-6 contributions and \(SU(3) \) flavor  breaking. 
At NLO, applying the same directional quadrature prescription, we predict that
\begin{equation}\label{4.6}
	{\cal R}_{cc}=0.70^{+0.05}_{-0.06}\,,
	\qquad
	{\cal R}_{cc}^{\Omega}=0.20^{+0.01}_{-0.03}\,,
\end{equation}
\begin{equation}
	{\cal R}_{bb}=0.10^{+0.01}_{-0.05}\,,
	\qquad
	{\cal R}_{bb}^{\Omega}=0.11^{+0.01}_{-0.05}\,.
\end{equation}
All four asymmetries are predicted to be positive. 
 Among them, ${\cal R}_{cc}$ is enormously large,
 reflecting  the strong $W$-exchange enhancement in
$\Xi_{cc}^{+}$. By contrast, the more moderate but still clearly nonzero values of
${\cal R}_{bb}$ and ${\cal R}_{bb}^{\Omega}$ indicate that the same mechanism remains operative
in the doubly bottom sector, albeit in a less dramatic form. 
 These decay width  asymmetries therefore provide  useful benchmark
observables for future experimental tests of HQE.

Another noteworthy aspect is the semileptonic sector~\cite{Shao:2025qwp}, collected in Table~\ref{tab:semi}.  Since each doubly heavy baryon contains two heavy quarks, its free‑quark semileptonic width is naturally expected to be about twice that of the corresponding singly heavy baryon.  In the charm sector, $\Xi_{cc}^{++}$ does not receive 
four-quark operator corrections in the semileptonic decays. 
Therefore, the simple free‑quark picture is expected to work particularly well for $\Xi_{cc}^{++}$, leading to the approximate relation $\Gamma(\Xi_{cc}^{++} \to X e^+) \approx 2\,\Gamma(\Lambda_c^+ \to X e^+)$.  Small deviations from this relation arise from differences in the bag radii, as well as from Cabibbo‑suppressed channels, which induce nonzero $\Gamma_{6}^{\rm SL} $ in  $\Lambda_c^+$.  Using the latest BESIII data, one finds $\Gamma(\Lambda_c^+ \to X e^+) = (1.31 \pm 0.04)\times 10^{-13}~\text{GeV}$~\cite{BESIII:2022cmg}, while our analysis yields $\Gamma(\Xi_{cc}^{++} \to X e^+) = (3.3 \pm 0.9)\times 10^{-13}~\text{GeV}$, where the dominant uncertainty comes from the variation of  $m_c$.  Conversely, if one takes the measured $\Gamma(\Lambda_c^+ \to X e^+)$ as an input, one obtains an effective value $m_c = 1.52~\text{GeV}$ at NLO, which still lies within the adopted range.  Overall, since dimension-6 contributions are absent, the semileptonic modes of $\Xi_{cc}^{++}$ provide an additional consistency check of our framework.  We also consider the saturation of the low‑lying states, defined by
\begin{equation}
    S_e = \frac{
\sum 
\Gamma(\Xi_{cc}^{++}  \to 
{\cal B}_c 
\,e^+ \nu_e) 
}{\Gamma(\Xi_{cc} ^{++} \to X e^+)}\,,
\end{equation}
where ${\cal B}_c$ runs through 
$  \Xi_c ^+ , \Xi_c^{\prime +} , \Lambda_c^+, \Sigma_c^+ 
$
Combining our inclusive result with the BM calculations of exclusive widths~\cite{Geng:2022uyy}, we find
$ 
S_e = (52^{+15}_{-14})\% .
$    In contrast, $\Lambda_c^+ \to \Lambda e^+ \nu_e$ 
and 
$\Lambda_c^+ \to n e^+ \nu_e$ 
makes up $(95 \pm 5)\%$ of $\Lambda_c^+ \to X e^+$~\cite{PDG2004}.

	\begin{table}
		\centering
		\caption{Semileptonic decay widths  given in units of $10^{-13}\,\text{GeV}$.
The uncertainties are obtained by combining the positive and negative variations separately in quadrature.}
	\label{tab:semi}
    \begin{center}
	\begin{tabular}{lccc}
		\hline
        \hline
		${\cal B}$ & $\Gamma({\cal B}\to e\,X)$ & $\Gamma({\cal B}\to \mu\,X)$ & $\Gamma({\cal B}\to \tau\,X)$ \\
		\hline
		$\Xi_{cc}^{++}$ & $3.32\pm0.86$ & $3.21\pm0.84$ &  \\
		$\Xi_{cc}^{+}$  & $3.34\pm0.86$ & $3.23\pm0.84$ &  \\
		$\Omega_{cc}^{+}$ & $2.01^{+1.00}_{-0.99}$ & $1.87^{+0.98}_{-0.97}$ &  \\
\hline
		$\Xi_{bb}^{0}$ & $0.91\pm0.13$ & $0.90\pm0.13$ & $0.22\pm0.05$ \\
		$\Xi_{bb}^{-}$ & $0.89^{+0.12}_{-0.11}$ & $0.88^{+0.12}_{-0.11}$ & $0.21\pm0.04$ \\
		$\Omega_{bb}^-$   & $0.90\pm0.11$ & $0.90\pm0.11$ & $0.21\pm0.04$ \\ 
		\hline
        \hline
	\end{tabular}
    \end{center}
\end{table}

\begin{figure}[t]
	\centering
	\includegraphics[width=0.78\textwidth]{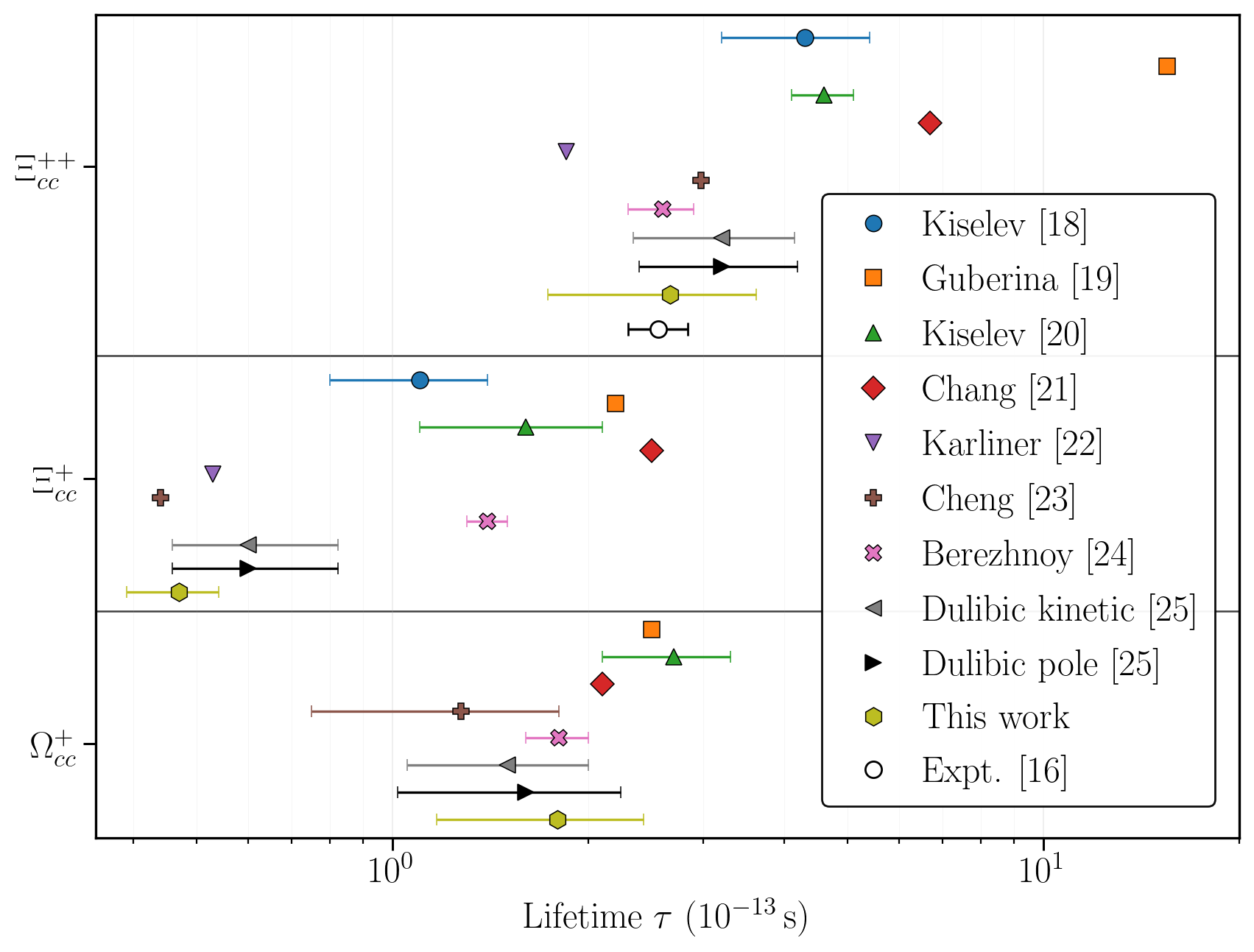}
	\caption{Theoretical predictions for the lifetimes of doubly charmed baryons, compared with the experimental value of $\tau(\Xi_{cc}^{++})$ from LHCb~\cite{LHCb:2018zpl}.  The plotted results are taken from Refs.~\cite{Kiselev:1999,Guberina,Kiselev:2002,Chang,Karliner:2014,Cheng:doubly,Berezhnoy,Dulibic:2023jeu}, together with our NLO result.  For Ref.~\cite{Berezhnoy}, the quoted uncertainties correspond to the analysis using $(m_c,m_s)=(1.73\pm0.07, 0.35\pm0.20)~\text{GeV}$ fitted to the LHCb measurement of $\tau(\Xi_{cc}^{++})$.}
	\label{fig:lifetime_dcc}
\end{figure}

We now compare our results with other works.  Figure~\ref{fig:lifetime_dcc} shows theoretical predictions for the lifetimes of doubly charmed baryons, arranged in chronological order except for the experimental value of $\tau(\Xi_{cc}^{++})$ from LHCb~\cite{LHCb:2018zpl}.  Our lifetime predictions are close to those of Ref.~\cite{Dulibic:2023jeu}, which also includes full NLO corrections except for the dimension-7 contributions, with only a slight difference in the central value of $\tau_{\Xi_{cc}^+}$; the two results remain consistent within uncertainties.  We note that $L^d _{\Xi_{cc}^+}$ is found to be $(-40 \pm 3 \pm 1)\times 10^{-3}~\text{GeV}^3$ in Ref.~\cite{Dulibic:2023jeu}, in good agreement with our result.  However, we have additionally taken into account the evolution of the matrix elements in eq.~\eqref{51}, which makes our predicted $\Xi_{cc}^+$ lifetime shorter than theirs.  A useful cross‑check is the total semileptonic width of $\Omega_{cc}$, which we find to be $3.88(1.94)(19)(+23)(+30)~10^{-13}$~GeV, whereas Ref.~\cite{Dulibic:2023jeu} yields $7.04 (3.55)^{+0.07}_{-0.66}~10^{-13}$~GeV 
using the pole mass scheme.  The main reason is that they used $\langle P_3^q \rangle_{\Omega_{cc}} = \Lambda_{\rm QCD}\,{\cal L}_{\Omega_{cc}}^s$ with $\Lambda_{\mathrm{QCD}} = 0.33$~GeV, while in the BM   $\Lambda_{\mathrm{QCD}}$ should be  the strange‑quark energy $E_s=0.50$~GeV instead.  The situation is similar to the discrepancy found for $\Omega_c^0$~\cite{Gratrex:2022xpm,Cheng:2023jpz}, and future tests of semileptonic decays should help resolve this issue.  From $\tau(\Xi_{cc}^+)/\tau(\Xi_{cc}^{++}) = 0.22 \pm 0.05^{+0.04}_{-0.03}$ and $\tau(\Omega_{cc}^+)/\tau(\Xi_{cc}^{++}) = 0.57 \pm 0.15^{+0.04}_{-0.03}$ in the pole mass scheme of Ref.~\cite{Dulibic:2023jeu}, we find ${\cal R}_{cc} = 0.64 \pm 0.07^{+0.04}_{-0.05}$ and ${\cal R}_{cc}^{\Omega} = 0.27 \pm 0.12^{+0.02}_{-0.03}$, which are in good agreement with our predicted values in eq.~\eqref{Rcc}.

\begin{figure}[t]
	\centering
	\includegraphics[width=0.7\textwidth]{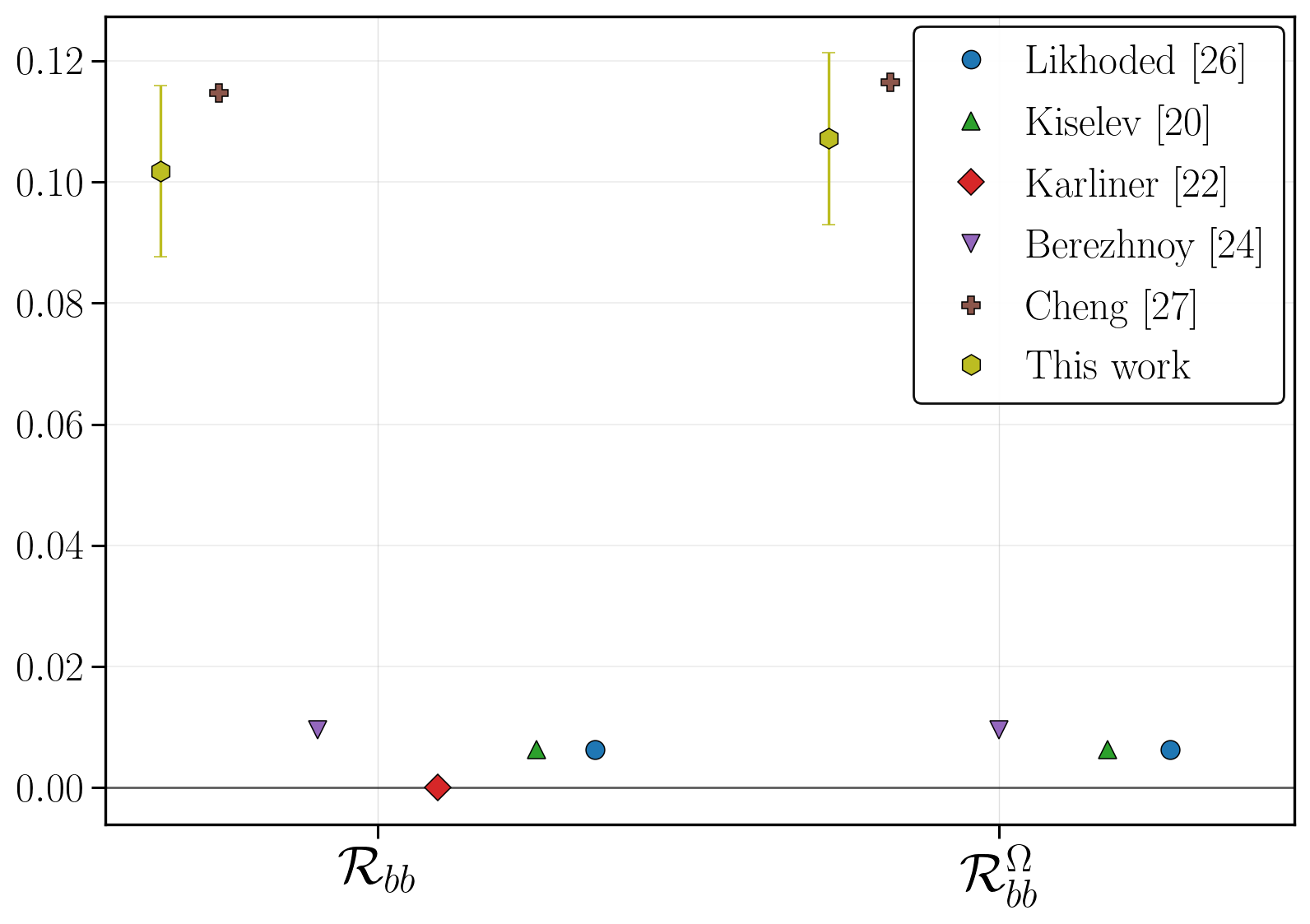}
	\caption{Comparison of the decay width asymmetries ${\cal R}_{bb}$ and ${\cal R}_{bb}^{(\Omega)}$ for doubly bottom baryons from different theoretical approaches, including our NLO result.  The literature values are taken from Refs.~\cite{Likhoded:1999yv,Kiselev:2002,Karliner:2014,Berezhnoy,Cheng:2019sxr}.    No uncertainties are shown for the other predictions as the correlations are unknown.}
	\label{fig:lifetime_dbb}
\end{figure}

For the bottom baryons, we compare ${\cal R}_{bb}$ and ${\cal R}_{bb}^{\Omega}$ with the literature in Fig.~\ref{fig:lifetime_dbb}.  We do not compare the lifetimes directly, because the bottom‑hadron lifetimes are dominated by free‑quark decays, and comparing the lifetimes mainly amounts to comparing the input value of $m_b$, which is plagued by renormalon ambiguity.  Except for an earlier work from one of the authors~\cite{Cheng:2019sxr}, most of the literature finds that  ${\cal R}_{bb}$ and 
${\cal R}_{bb}^\Omega$
 are negligible.  However, as we have stressed, the $W$‑exchange diagram plays a very important role in heavy‑baryon decays, even for bottom baryons.  The correction is around $15\%$ in $\Xi_{bb}^0$, generating the splitting in ${\cal B}_{bb}$ lifetimes.   

\section{Conclusion}
 In this work, we have studied the lifetimes and inclusive semileptonic decay widths of doubly heavy baryons within the framework of HQE. Our analysis includes the next-to-leading-order contributions to the dimension-3, -5, and -6 operators, together with the leading dimension-7 corrections, while all relevant charm- and bottom-quark decay channels are taken into account. The nonperturbative matrix elements are evaluated in the BM with translationally improved baryon wave functions, and their scale dependence is incorporated through the evolution from the hadronic scale to the heavy-quark scale.
 
 Our results show that HQE works much better for the doubly bottom baryons than for the doubly charmed ones. In the bottom sector, the dimension-3 terms remain dominant and the higher-order contributions behave as genuine corrections, indicating a reasonably convergent expansion. By contrast, in the charm sector the dimension-6 spectator effects are strongly enhanced and can even compete with or exceed the leading contributions in some channels. Numerically, we obtain
$
	 ( \tau_{\Xi_{cc}^{++}} , \tau_{\Xi_{cc}^{+}} , \tau_{\Omega_{cc}^{+}} )
	 = ( 2.67^{+0.95}_{-0.94},\, 0.47^{+0.09}_{-0.08},\, 1.79^{+0.64}_{-0.62} )
	 \times 10^{-13}\,{\rm s}
$
and
$
	 ( \tau_{\Xi_{bb}^{0}} , \tau_{\Xi_{bb}^{-}} , \tau_{\Omega_{bb}^{-}} )
	 = ( 0.75^{+0.13}_{-0.11},\, 0.92^{+0.15}_{-0.15},\, 0.93^{+0.15}_{-0.15} )
	 \times 10^{-12}\,{\rm s}.
$ Because of the slow convergence of HQE for charm, these charmed-baryon lifetimes, most notably \(\tau( \Xi_{cc}^{+}) \), should be taken as indicative estimates. {The quoted errors summarize the parameter, radius, and scheme variations described above, but not a separate HQE-truncation uncertainty.} In particular, the $W$-exchange topology plays a crucial role in generating the large lifetime splitting in the doubly charmed sector and remains phenomenologically important even in the doubly bottom sector.
 
 We have further shown that the decay-width asymmetries provide clean probes of spectator effects. In particular, ${\cal R}_{cc}$ is predicted to be large due to the strong $W$-exchange contribution in $\Xi_{cc}^{+}$, whereas ${\cal R}_{bb}$ and ${\cal R}_{bb}^{\Omega}$ are both of order ten percent. Since these observables are less sensitive to the overall heavy-quark-mass normalization, they provide useful benchmarks for comparing different theoretical approaches. We have also separated the nonleptonic and semileptonic contributions, allowing the pattern of spectator effects to be traced channel by channel. In the semileptonic sector, the free-quark picture works particularly well for $\Xi_{cc}^{++}$, for which $\Gamma(\Xi_{cc}^{++}\to X e^+) \approx 2\,\Gamma(\Lambda_c^+\to X e^+)$ is approximately satisfied, while dimension-6 spectator corrections remain important especially for $\Omega_{cc}^{+}$.
We also examine the saturation by low-lying states, found to be  $S_e=(52^{+15}_{-14})\% $. By comparison, $\Lambda_c^+ \to \Lambda e^+ \nu_e$ accounts for $(95\pm5 )\%$ of $\Lambda_c^+ \to X e^+$, indicating much weaker low-lying-state saturation in the $\Xi_{cc}$ sector.

	\begin{acknowledgments}
		This research was supported in part by the Ministry of Science and Technology of R.O.C. under Grant No. MOST-114-2112-M-001-039 and 
		the National Natural Science Foundation of China
		under Grant No. 12575096.
	\end{acknowledgments}

\end{document}